\documentclass{article} 

\usepackage[utf8]{inputenc} 
\usepackage{verbatim}
\usepackage{calc}
\usepackage[margin=1.2in]{geometry}
\usepackage{amssymb}
\usepackage{amsmath}
\usepackage{hhline} 
\usepackage{xcolor}
\usepackage{graphicx}
\usepackage{lipsum}
\usepackage{eso-pic} 
\usepackage[toc,page]{appendix}
\usepackage{tikz} 
\usepackage{lscape} 
\usepackage{hyperref}
\usepackage{epigraph}
\usepackage{subcaption}
\usepackage{units}
\usepackage{float}
\usepackage{wrapfig}
\usepackage{dsfont}
\usepackage{xfrac}
\usepackage{caption}
\usepackage{pgfplots}
\usepackage{filecontents}
\usetikzlibrary{automata}
\usetikzlibrary{arrows,petri,topaths,decorations}
\usetikzlibrary{decorations.pathreplacing,shapes.arrows,shapes.geometric,decorations.markings}
\usetikzlibrary{decorations.pathmorphing,backgrounds,positioning,fit,petri,calc}
\usepackage{tkz-berge}
\usepackage{mathtools}
\usepackage{tikz-cd}
\usepackage{xcolor}
\usepackage{multirow, tabularx}
\usepackage{makeidx} 
\usepackage{complexity} 

\definecolor{linkcolour}{rgb}{0.67, 0.12, 0.18}
\hypersetup{ 
    colorlinks,
    linkcolor={linkcolour},
    citecolor={linkcolour},
    urlcolor={linkcolour}
}

\definecolor{babyblue}{rgb}{0.5, 0.12, 0.18}

\makeatletter
\newcommand{\vast}{\bBigg@{4}}
\newcommand{\Vast}{\bBigg@{5}}
\makeatother

\input{Qcircuit}

\usepackage{fancyhdr}
 
\definecolor{headerlinecolour}{rgb}{0.75,0.73, 0.65} 
\renewcommand{\headrulewidth}{1.5pt}
\renewcommand{\headrule}{\hbox to\headwidth{%
  \color{headerlinecolour}\leaders\hrule height \headrulewidth\hfill}}

\newcounter{arraycard}
\def\arrayLength#1{%
  \setcounter{arraycard}{0}%
  \foreach \x in #1{%
    \stepcounter{arraycard}%
  }%
  \the\value{arraycard}%
}

\begin{document}

\begingroup
\thispagestyle{empty}
\vspace*{2cm}
\centering
\par\normalfont\fontsize{20}{20}\sffamily\selectfont
\vspace*{0.7cm}
\textbf{Advances in quantum machine learning}\\
{\LARGE }\par 
\vspace*{1cm}
{\large{J. C. Adcock, E. Allen, M. Day*, S. Frick, J. Hinchliff,  M. Johnson,\\ S. Morley-Short,  S. Pallister, A. B. Price, S. Stanisic } }\par 
\vspace*{0.5cm}
\small{\today} \par
\vspace*{0.7cm}
{\small{Quantum Engineering Centre for Doctoral Training, University of Bristol, UK}}\par
\vspace*{0.7cm}
{\small{*Corresponding author: matt.day@bristol.ac.uk}}\par
\vspace*{0.7cm}
\endgroup
\newpage
\epigraph{\textit{``We can only see a short distance ahead, but we can see plenty there that needs to be done."}}{\textit{Alan Turing ``Computing machinery and intelligence." Mind (1950): 433-460.}}

\epigraph{\textit{``The question of whether machines can think... is about as relevant as the question of whether submarines can swim.''}}{\textit{Edsger W. Dijkstra  ``The threats to computing science." (1984).}}

\section*{Preface}
\addcontentsline{toc}{section}{Preface}
\thispagestyle{empty}

Created by the students of the Quantum Engineering Centre for Doctoral Training, this document summarises the output of the `Cohort Project' postgraduate unit. The unit's aim is to get all first year students to investigate an area of quantum technologies, with a scope broader than typical research projects. The area of choice for the 2014/15 academic year was quantum machine learning. 

The following document offers a hybrid discussion, both reviewing the current field and suggesting directions for further research. It is structured such that Sections~\ref{sec:CML} and~\ref{sec:LTable} briefly introduce classical machine learning and highlight useful concepts also relevant to quantum machine learning. Sections~\ref{sec:QMLAlgorithms} and~\ref{sec:Implementations} then examine previous research in quantum machine learning algorithms and implementations, addressing algorithms' underlying principles and problems. Section~\ref{sec:Challenges} subsequently outlines challenges specifically facing quantum machine learning (as opposed to quantum computation in general). Section~\ref{sec:conclusion} concludes by identifying areas for future research and particular problems to be overcome.

The conclusions of the document are as follows. The field's outlook is generally positive, showing significant promise (see, for example, Sections \ref{sec:DL}, \ref{sec:HHLalg} and \ref{sec:kNN}). There is, however, a body of published work which lacks formal derivation, making it difficult to assess the advantage of some algorithms. Significant real-world applications demonstrate an obvious market for implementing learning algorithms on a quantum machine. We have found no reason as to why this cannot occur, however there are appreciable hurdles to overcome before one can claim that it is a primary application of quantum computation. We believe current focus should be on the development of quantum machine learning algorithms that show an appreciation of the practical difficulties of implementation.

A table summarising the advantages of the algorithms discussed in this document can be found in Appendix~\ref{app:tablealgorithms}. 

\newpage
\section*{List of Acronyms}
\addcontentsline{toc}{section}{List of Acronyms}
\thispagestyle{empty}
\begin{description}
\item[ANN:] Artificial neural network
\item[BM:] Boltzmann machine
\item[BN:] Bayesian network
\item[CDL:] Classical deep learning
\item[CML:] Classical machine learning
\item[HMM:] Hidden Markov model
\item[HQMM:] Hidden quantum Markov model
\item[$k$-NN:] $k$-nearest neighbours
\item[NMR:] Nuclear magnetic resonance
\item[PCA:] Principal component analysis
\item[QBN:] Quantum Bayesian network
\item[QDL:] Quantum deep learning
\item[QML:] Quantum machine learning
\item[QPCA:] Quantum principal component analysis
\item[QRAM:] Quantum random access memory
\item[RAM:] Random access memory
\item[SQW:] Stochastic quantum walk
\item[SVM:] Support vector machine
\item[WNN:] Weightless neural network

\end{description}
\newpage
\thispagestyle{empty}
\tableofcontents
\newpage

\section{Introduction} \label{sec:Introduction}
Machine learning algorithms are tasked with extracting meaningful information and making predictions about data. In contrast to other techniques, these algorithms construct and/or update their predictive model based on input data. The applications of the field are broad, ranging from spam filtering to image recognition, demonstrating a large market and wide societal impact~\cite{Witten2005}.

In recent years, there have been a number of advances in the field of quantum information showing that particular quantum algorithms can offer a speedup over their classical counterparts~\cite{Algorithmzoo}. It has been speculated that application of these techniques to the field of machine learning may produce similar results. Such an outcome would be a great boost to the developing field of quantum computation, and may eventually offer new practical solutions for current machine learning problems. For a general introduction to the field of quantum computation and information, see reference~\cite{Nielsen2011}.

\subsection{Classical Machine Learning}\label{sec:CML}
Machine learning algorithms are almost exclusively used to categorise instances of data into classes that can either be user defined, or found from the intrinsic structure of the data. Consider the dataset $\boldsymbol{X} = \{ \boldsymbol{x_1},\boldsymbol{x_2},...,\boldsymbol{x_n} \}$ where each $\boldsymbol{x_i}$ is a data point that is itself defined by a number of parameters $\boldsymbol{x_i} = (x_i^1,x_i^2,...,x_i^m)$. An example dataset would be a collection of images, where each image is defined by parameters such as the number of pixels it contains or the colour content across a certain region. The classification of this data could be to sort between images that contain cars and those that do not. It is the role of the machine learning algorithm to acquire the classification rule.

Broadly speaking, machine learning algorithms can be broken into three types: supervised, unsupervised and reinforced learning. Supervised learning is based on having a predefined set of training data $\boldsymbol{X_T}$ (known as a `labelled' dataset) which contains data points that have already been correctly classified to produce a set of classifications $\boldsymbol{Y}= \{ y_1,y_2,...,y_n \}$, where $y_i$ is the classification of the data point $\boldsymbol{x_i}$. The machine learning algorithm takes in $\boldsymbol{Y}$ and $\boldsymbol{X_T}$ and optimises internal parameters until the closest classification of the training set to $\boldsymbol{Y}$ has been reached. Once the machine has learned, it is then fed new, unlabelled data $\boldsymbol{X}$ which it classifies but does no learning from. In contrast, reinforced learning has no training set. Instead, the user dynamically inputs the result of the machine classification on an unmarked dataset as either correct or incorrect. This is used to feed back through the algorithm/machine and results in a learning process. Cases where the classification sets are not predefined (i.e. where  $\{ \boldsymbol{Y} \}$ does not exist) come under the bracket of unsupervised learning; this happens typically because the datasets are too large or complex. Here, the machine looks for any natural structure within the data. For example, consider a very large and complex dataset ($n,m \gg 1$). Each data point of this set is a vector in a high-dimensional data space. The complexity of the data may make it impossible for the user to predefine a desired output or method of learning, which makes supervised or reinforced learning difficult. However, an unsupervised clustering algorithm such as $k$-means (Section~\ref{sec:kmeans_lloyd}) can be used, which may split the data into distinct clusters. These clusters can give information on the relationships between different features of the data and can be used to later classify new instances of data. An application of this analysis is market research, where data can be the results of a market survey. The goal of the user is to segment the market into consumers that have similar attributes, which can therefore be exposed to similar marketing strategies~\cite{Wedel2012}. 

\subsection{Quantum Machine Learning}\label{sec:LTable}
\begin{figure}[hbtp]
\centering
\includegraphics[scale=0.7]{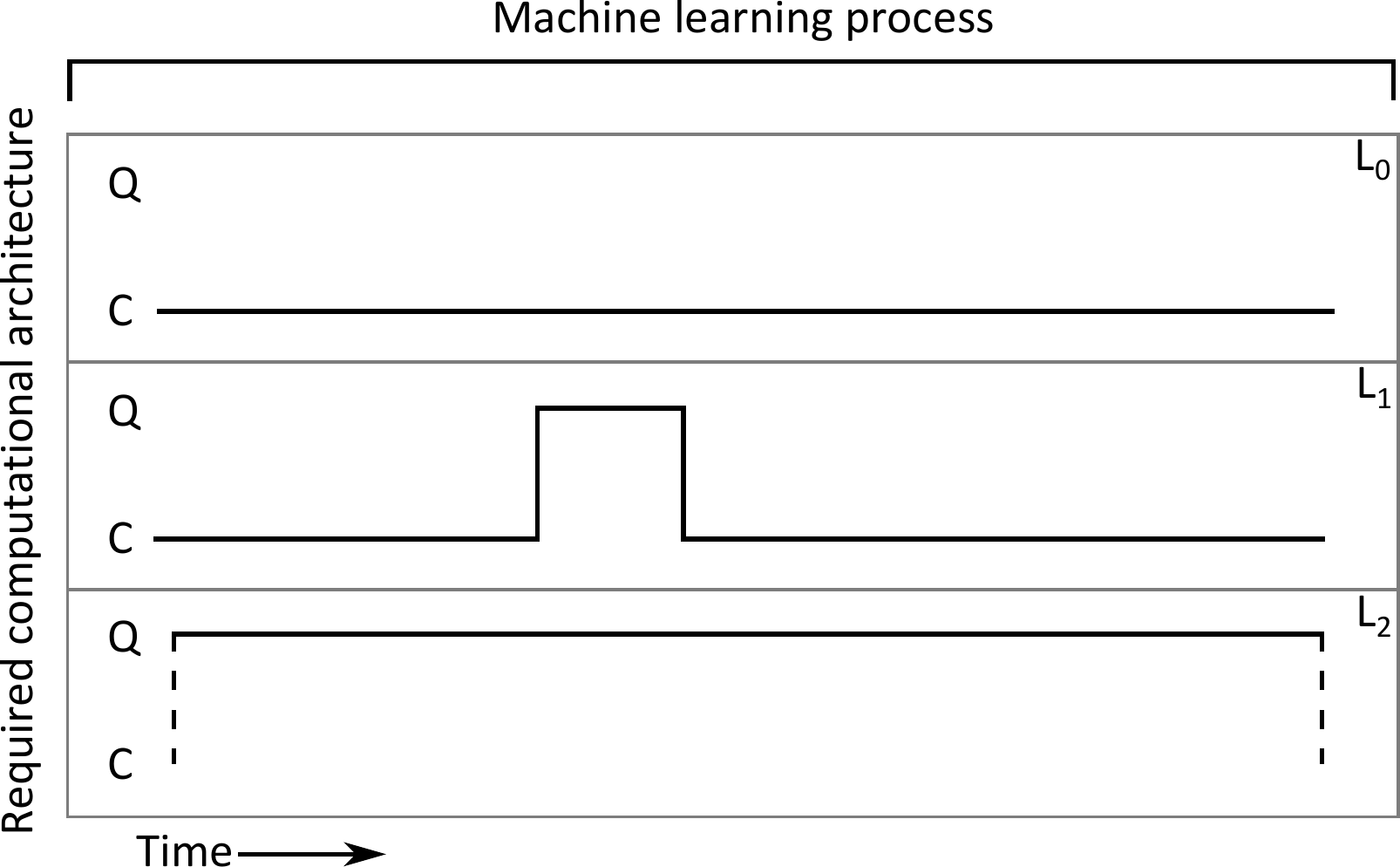}
\caption{Pictorial representation of the learning categories. Here, ``Q" and ``C" denote quantum and classical computation respectively. The solid line represents the minimal computational requirements of the protocol during the learning process. The dashed lines in $L_2$ represent the possibility of having data that does not need to be represented classically. \label{fig:QMLcategories}}
\end{figure}

The first problem encountered with quantum machine learning (QML) is its definition. By considering a number of scenarios, we aim to clarify how machine learning and quantum mechanics can be combined, and whether \emph{we} will consider these to be QML. This is subjective and any deep theoretical meaning requires further clarification. The considered scenarios have become a useful tool when considering what form a QML protocol could take and how it could be implemented practically. We will group the scenarios into categories of learning, which are loosely based on definitions made by A\"\i meur, Brassard, and Gambs~\cite{Aimeur2006,Gambs2008}. Figure~\ref{fig:QMLcategories} is a pictorial representation of the categories, which the reader should consult as they are defined.

First, we consider the classical case, which defines the category $L_0$. Traditional classical machine learning (CML) is included here, where a classical machine learns on classical data. Also included is machine learning performed on a classical machine but with data that has come from a quantum system; this is beginning to be applied within a number of schemes~\cite{Hentschel2010}. The second category, $L_1$, we will denote as QML. Here, the learning process can predominantly be run with classical computation, with only part of the protocol (e.g. a particular sub-routine) requiring access to a quantum computer. That is, the speedup gained by quantum computation comes directly from a \emph{part} of the process (for example by using Grover’s~\cite{Grover1997} or Shor’s~\cite{Shor1997} algorithm). As with $L_0$, we will include in $L_1$, cases where classical data fed to the algorithm originates from both classical and quantum systems. We note here that although only a small part of the algorithm requires a quantum computer, the overall process suffers no drawback from being computed entirely on a quantum machine, as classical computation can be performed efficiently on a quantum computer~\cite{Nielsen2011,Toffoli1980}. However, it may be more practical to use a quantum-classical hybrid computational system. In either case, protocols within this category will have to carefully consider any limitations from read-in and read-out of the data on either side of the quantum computational process.

We will denote the final learning category as $L_2$; this is also considered to be QML. Here, the algorithm contains no sub-routine that can be performed equally well on a classical computer. Like before, the input and output data can be classical. However, one can imagine cases where the data naturally takes quantum form\footnote{An example could be learning the noise model of a quantum channel. One could send quantum states through the channel and pass these directly to a learning machine. The QML algorithm then learns on the noise and classifies any new states passed through the channel. Once classified, the states could be corrected by another method. This process may not require any classical interface at all.}.
\newpage

\subsection{Comparison of Machine Learning Algorithms} \label{sec:Comparison}

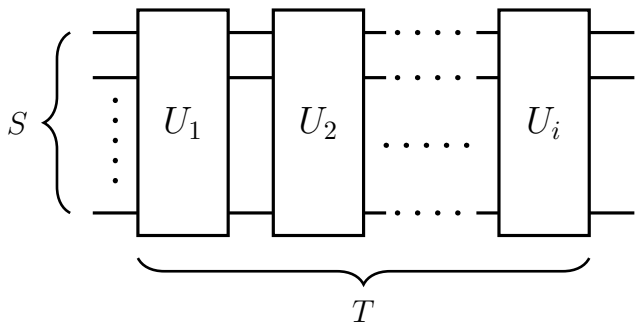
\begin{figure}[h]
		\begin{tikzpicture}[scale=0.6]
                    	\draw[very thick] (0,3) -- (6.5,3);
                    	\draw[very thick] (8.5,3) -- (12,3);
                    	\draw[very thick] (0,2) -- (6.5,2);
                    	\draw[very thick] (8.5,2) -- (12,2);
                    	\draw[very thick] (0,-1) -- (6.5,-1);
                    	\draw[very thick] (8.5,-1) -- (12,-1);
                    	\draw[fill=white,very thick] (1,3.5) rectangle (3,-1.5);
                    	\node at (2,1) {\Large$U_1$};
                    	\draw[fill=white,very thick] (4,3.5) rectangle (6,-1.5);
                    	\node at (5,1) {\Large$U_2$};
                    	\draw[fill=white,very thick] (9,3.5) rectangle (11,-1.5);
                    	\node at (10,1) {\Large $U_i $};
                    	\draw[line width=2pt, line cap=round, dash pattern = on 0pt off 7.5] (0.5,1.5) -- (0.5,-0.5);
                    	\draw[line width=2pt, line cap=round, dash pattern = on 0pt off 7.5] (6.75,3) -- (8.25,3);
                    	\draw[line width=2pt, line cap=round, dash pattern = on 0pt off 7.5] (6.75,2) -- (8.25,2);
                    	\draw[line width=2pt, line cap=round, dash pattern = on 0pt off 7.5] (6.5,0.5) -- (8.5,0.5);
                    	\draw[line width=2pt, line cap=round, dash pattern = on 0pt off 7.5] (6.75,-1) -- (8.25,-1);
                    	\draw [line width=1pt,decorate,decoration={brace,mirror,amplitude=10pt},xshift=-0pt,yshift=0pt]
	(-0.5,3) -- (-0.5,-1) node [black,midway,xshift=-20pt] {\large $S$};
			\draw [line width=1pt,decorate,decoration={brace,mirror,amplitude=10pt},xshift=0pt,yshift=0pt] (1,-2) -- (11,-2) node [black,midway,yshift=-20pt] {\large $T$};
		\end{tikzpicture}
\centering
\caption{The roles of space $S$ and time $T$ in the circuit model for quantum computation.}
\label{fig:resources}
	\end{figure}

In order to understand the potential benefits of QML, it must be possible to make comparisons between classical and quantum machine learning algorithms, in terms of speed and classifier performance. To compare algorithms, computer scientists consider two characteristic resources:


\begin{itemize}

\item{\bf{Space}, $S$:} The amount of computational space needed to run the algorithm. Formally, `space' refers to the number of qubits required. For $S$ qubits, the dimension of the relevant Hilbert space is $2^S$. It is important to distinguish between these two quantities, as there is an exponential factor between them. 

\item{\bf{Time}, $T$:} The time taken to train and then classify within a specified error. Formally, `time' refers to the number of operations required and, in the quantum circuit model, can be expressed as the number of consecutive gates applied to the qubits.

\end{itemize}

Figure~\ref{fig:resources} shows how time and space are represented in quantum circuit diagrams. These are typically functions of the following variables:
\begin{itemize}

\item{\bf{Size of training dataset}, $n$:} The number of data points in the training set supplied to an algorithm. 

\item{\bf{Size of input dataset}, $N$:} The number of data points to be classified by an algorithm.

\item{\bf{Dimension of data points}, $m$:} The number of parameters for each data point. In machine learning, each data point is often treated as a vector, where the numeric value associated with each feature is represented as a component of the vector. 

\item{\bf{Error}, $\epsilon$:} The fraction of incorrect non-training classifications made by the algorithm.

\end{itemize}

\begin{figure}[h]
	\centering
	\includegraphics[width=0.5\textwidth]{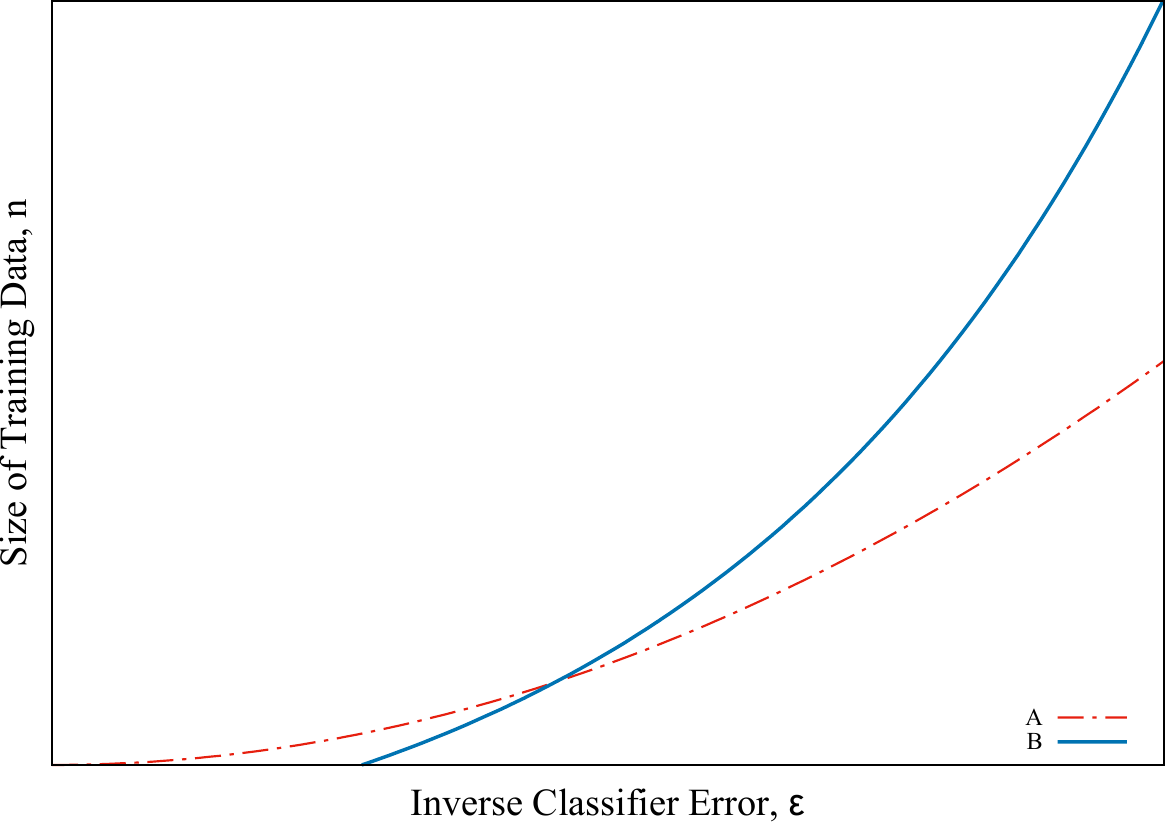}
\caption{Example plot illustrating the inverse classifier error scalings with increasing size of the training dataset, for two different algorithms.}
\label{fig:scalinggraph}
\end{figure}

Note that not all algorithms necessarily have resource scalings dependent on all the above variables. For example, unsupervised learning does not depend on $n$, as no training data exists. Figure~\ref{fig:scalinggraph} depicts the scaling of two hypothetical algorithms with error, both exhibiting differing convergence properties. This aims to highlight the situational nature of algorithm comparison. Here, the assertion of which algorithm is ``better'' depends entirely on what level of error one is willing to accept. It is also important to note that not all algorithms will converge to $\epsilon = 0$ given infinite resources.

\section{Quantum Machine Learning Algorithms}\label{sec:QMLAlgorithms}

In this section, we outline several attempts at quantum machine learning algorithms. We begin by exploring neural networks, before moving onto clustering based algorithms, and ending with several other notable attempts. In order to give a complete overview of the field, we have not outlined any algorithms in detail, and the interested reader should follow the relevant references. For other reviews in quantum machine learning, see references \cite{Wittek2014} and \cite{Schuld2014d}.

\subsection{Neural Networks}\label{sec:NN}
\begin{figure}[h]
	\centering
	\includegraphics[width=\textwidth]{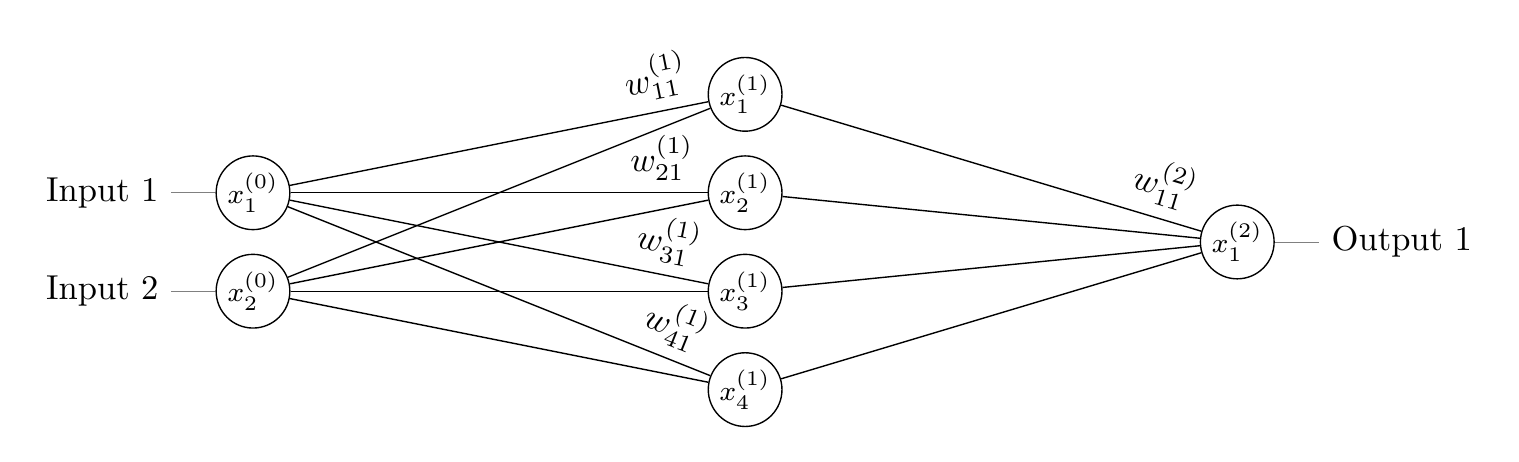}
\caption{
A feed-forward neural network.
Input nodes are on the left, the number of which depends on how many parameters a single training data point has.
If the dimension of the input vector is $m = 2$, there will be $2$ input nodes - we are currently ignoring an extra node that is usually added to the input for bias purposes.
The last layer on the right is the output layer, which tells us how the input has been classified.
Here we see only one output node, making this a binary classifier.
The layers in between the input and the output are called hidden layers, and there can be as many of them as the user finds suitable for the dataset being classified.
Similarly, the number of nodes per hidden layer can be varied as much as necessary.
Each of the nodes represents a neuron and the weight associated with an inter-node edge marks the ``strength'' of the link between the neurons.
The higher the weight between two nodes, the greater the influence on the linked node.
} \label{fig:NN}
\end{figure}

Classical artificial neural networks (ANN) are based on graphs which usually are layered. The nodes of the graph correspond to neurons, and the links to synapses (see Figure \ref{fig:NN}).
They can be used for supervised or reinforcement learning.
Just like neurons, the nodes have an activation threshold, which is commonly described by a sigmoid function.
Each edge in the graph is associated with a weight, making some neurons more relevant in the activation of their neighbours.

A perceptron is a feed-forward single-layer ANN containing only an input and binary output layer.
In a feed-forward layered ANN (Figure \ref{fig:NN}), the output of each layer generates the input to the next one.
For this purpose, we denote input to the nodes in layer $k$ as $\boldsymbol{x}^{(k)}_i$, and the output as $\boldsymbol{y}^{(k)}_i$.
For example, given input vector $ \boldsymbol{x}^{(0)}_i := \boldsymbol{x}_i = (x_i^1, x_i^2, ..., x_i^m) =: \boldsymbol{y}^{(0)}_i$, the input to the second node in the first layer is then calculated as $x^{2, (1)}_i = \sum_{t = 1}^{m} w^{(1)}_{2,t} x_i^t$, and the output is $y^{2, (1)}_i = f(x^{(1), 2}_i)$ where $f$ is an ``activation function'', commonly a sigmoid.
If $f$ is a linear function, the multi-layer network is equivalent to a single layer network, which can only solve linearly separable problems.
We can see that input into layer $k$ is $\boldsymbol{x}^{(k)}_i = \boldsymbol{w}^{(k)} \cdot \boldsymbol{y}^{(k-1)}_i$, where $\boldsymbol{w}^{(k)}$ is a matrix of weights for layer $k$, quantifying to what degree different nodes in layer $k-1$ affect the nodes in layer $k$).
Training in this type of ANN is usually done through weight adjustment, using gradient descent and backpropagation.
The idea is that the output layer weights are changed to minimize the error, which is the difference between the calculated and the known output.
This error is then also backpropagated all the way to the first layer, adjusting the weights by gradient descent.

Hopfield networks are another ANN design, which are bi-directional fully-connected graphs, not arranged in layers \cite{Hopfield1982}.
Their connectivity means they can be used as associative memory, allowing retrieval of a complete pattern, given an incomplete input.
Hopfield networks retrieve this foreknown pattern as a minimum of its energy function (see Equation (\ref{eq:hop_net})) with high probability.
Other models also exist although, with the exception of weightless ANNs \cite{Aleksander2009},  many have been studied to a far lesser extent in the context of QML (for more information on classical ANN models, see reference \cite{Cheng1994}).

The linear algebraic structure of ANN computation means that, by choosing appropriate operators and states, quantum mechanics could be useful \cite{Altaisky2014}.
Specifically, some of the ANN steps require vector and matrix multiplication which can be intrinsic to quantum mechanics (although result retrieval may be problematic \cite{Aaronson2015}).

The earliest proposals of quantum ANNs date back to 1996 by Behrman et al. \cite{Behrman1996} and Toth et al. \cite{Toth1996}.
Upon further inspection, much of the work done in this area seems to be quantum inspired - the algorithms can be described, understood and run efficiently on classical devices, but the inception of their idea stems from quantum concepts.
For a comprehensive review, see Manju and Nigam \cite{Manju2012}.
As an example, the work of Toth et al. was based on quantum-dot cellular ANNs, but inter-cell communications were classical \cite{Toth1996}.
Ventura et al. contributed to the area in 1997, with papers on both quantum ANNs and associative memories.
Unfortunately, this work seemed to be mostly quantum-inspired \cite{Ventura2000}, incomplete \cite{Howell2000a} or contained misleading quantum notation \cite{Ricks2004}.
%
%
%

Behrman et al. have a large body of work which aims to show, by numerical simulations, that a quantum dot molecule can be used as a neural network \cite{Behrman1996,Behrman2000, Behrman2002, Behrman1999, Behrman2013, Behrman2008}.
These attempts are interesting as they try to use intrinsic properties of the system to construct an ANN.
However in their attempts (such as that detailed in reference \cite{Behrman2000}), several problems exist.
Firstly, the parameters are unphysical, i.e. a 65.8ps evolution time is shorter than the spontaneous emission rate of a quantum dot.
Also, the measurement modelling accepts error rates of up to 50\%, and the learning update rules are classical.

More recent efforts by De Oliveria, Da Silva et al. \cite{DeOliveira2008, Silva2010, DaSilva2012, DaSilva2012a} explore the idea of a quantum weightless neural network (WNN).
A WNN consists of a network of random access memory (RAM) nodes, and can take a number of different forms \cite{Aleksander2009}. This was one of the first machine learning implementations due to the ready availability of RAM in the 1970s, and the relative ease with which training could take place.
They develop a few approaches to a quantum version of the nodes used in WNN (quantum probabilistic logic node \cite{DeOliveira2008}, superposition based learning algorithm \cite{DaSilva2012a}, probabilistic quantum memory \cite{Silva2010}, $\ket{\psi}$-node \cite{DaSilva2012}), but they do not give a thorough analysis to compare these with the classical version, and no simulation is offered to strengthen the claims.
If quantum random access memory (QRAM) becomes available before other machine learning algorithms can be implemented, this area will be worth re-visiting.
Similarly,  recent work of Schuld et al. \cite{Schuld2014a, Schuld2014b, Schuld2015} seems to show promise.
However, we believe there are still limitations with elements of these schemes, and the existence of a speedup over classical algorithms (for a detailed analysis, see Appendix \ref{app:perceptron}).

A common approach to ANNs describes input vectors as quantum states, and weights as operators \cite{Altaisky2014, Zhou2007, Schuld2015}.
This approach, sometimes combined with measurement on each iteration of the algorithm to feed back the output classically, usually results in the algorithm being incomplete (missing an appropriate training step, for example \cite{Altaisky2014, Schuld2015}) or even disputed as incorrect \cite{DaSilva2014}. 
If the algorithm assumes that the measured quantum data can be reused as the original state, problems may arise. 

In relation to Hopfield networks and QRAM, more serious work has been done on quantum associative memory \cite{Trugenberger2001}.

We will now proceed to mention a few promising approaches for ANNs in more detail.
For an in-depth review, see Schuld et al. \cite{Schuld2014c}.

\subsubsection{Quantum Walks} \label{sec:QWalks}
Schuld et al. \cite{Schuld2014a} propose using quantum walks to construct a quantum ANN algorithm, specifically with an eye to demonstrate associative memory capabilities.
This is a sensible idea, as both discrete-time and continuous-time quantum walks are universal for quantum computation \cite{Childs2009, Lovett2010}.
In associative memories, a previously-seen complete input is retrieved upon presentation of an incomplete or noisy input.

The quantum walker position represents the pattern of the “active” neurons (the firing pattern).
That is, on an $n$-dimensional hypercube, if the walker is in a specific corner labelled with an $n$-bit string, then this string will have $n$ corresponding neurons, each of which is ``active'' if the corresponding bit is $1$.
In a Hopfield network for a given input state $\boldsymbol{x}$, the outputs are the minima of the energy function
\begin{equation}
E(x^1,...,x^n) = - \frac{1}{2} \sum_{i=1}^{n}\sum_{j=1}^{n} w_{ij} x^i x^j + \sum_{i=1}^{n} \theta_i x^i,
\label{eq:hop_net}
\end{equation}
where $x^i$ is the state of the $i$-th neuron, $w_{ij}$ is the strength of the inter-neuron link and $\theta_i$ is the activation threshold.
Their idea is to construct a quantum walker such that one of these minima (dynamic attractor state) is the desired final state with high probability.

The paper examines two different approaches.
First is the na\"{i}ve case, where activation of a Hopfield network neuron is done using a biased coin.
However they prove that this cannot work as the required neuron updating process is not unitary.
Instead, a non-linearity is introduced through stochastic quantum walks (SQW) on a hypercube.
To inject attractors in the walker's hypercube graph, they remove all edges leading to/from the corners which represent them.
This means that the coherent part of the walk can't reach/leave these states, thus they become sink states of the graph.
The decoherent part, represented by jump operators, adds paths leading to the sinks.
A few successful simulations were run, illustrating the possibility of building an associative memory using SQW, and showing that the walker ends up in the sink in a time dependent on the decoherent dynamics.
This might be a result in the right direction, but it is not a definitive answer to the ANN problem since Schuld et al. only demonstrate some associative memory properties of the walk.
Their suggestion for further work is to explore open quantum walks for training feed-forward ANNs.

\subsubsection{Deep Learning} \label{sec:DL}

\begin{figure}[h]
	\centering
	\includegraphics[width=\textwidth]{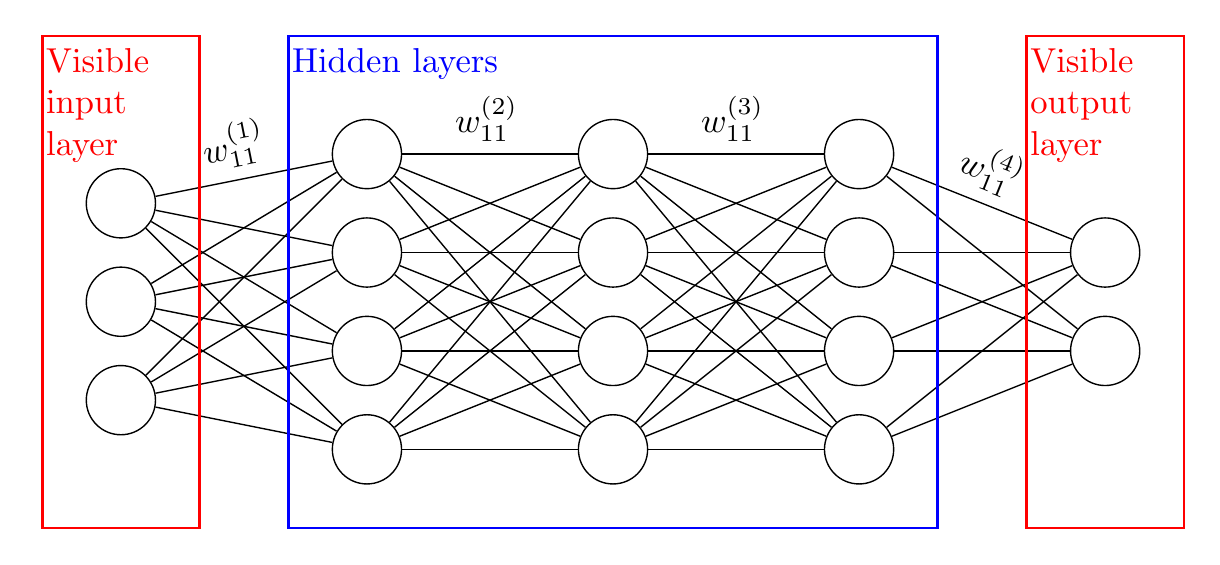}
\caption{
An example of a deep learning neural network with 3 hidden layers. For a Boltzmann machine, each layer is specified as a vector of binary components, with the edges between the vectors defined as a matrix of weight values. The configuration space of the graph is given by a Gibbs distribution with an Ising-spin Hamiltonian.
} \label{fig:Dl}
\end{figure}

The field of classical deep learning (CDL) has revolutionised CML \cite{LeCun2015}. Utilising multi-layer, complex neural networks, a deep learning algorithm can construct many layers of abstraction from large datasets. For example, given a large selection of car photos, a deep learning algorithm could first learn to classify the shape of a car, then the concept of `left' and `right' facing cars, and even progress to further details. These layers of abstraction have become a powerful resource in the machine learning community.
 
One class of deep learning networks is the Boltzmann machine (BM), for which the configuration of graph nodes and connections is given by a Gibbs distribution \cite{Ackley1985} 
\begin{equation}
P(\mathbf{v},\mathbf{h})=\frac{e^{-E(\mathbf{v},\mathbf{h})}}{\sum_{\mathbf{v},\mathbf{h}} e^{-E(\mathbf{v},\mathbf{h})}}~,
\end{equation}
where $\mathbf{v}$ and $\mathbf{h}$ are vectors of binary components, specifying the values of the visible and hidden nodes (see Section \ref{sec:NN} and Figure \ref{fig:Dl} for more details).
The energy of a given configuration resembles an Ising model at thermal equilibrium with weight values determining the interaction strength between nodes 
\begin{equation}
E(\mathbf{v},\mathbf{h})=-\sum_i a_i v_i - \sum_j b_j h_j - \sum_{i,j} w_{ij} v_i h_j~.
\end{equation}
The objective is then to minimise the maximum-likelihood of the distribution (the agreement of the model with the training data) using gradient descent. Classically, computing the required gradient takes time exponential in the number of nodes of the network and so approximations are made such that the BM configuration can be efficiently calculated \cite{Ackley1985}.
 
Recently, Wiebe et al. have developed two quantum algorithms that can efficiently calculate the BM configuration without approximations, by efficiently preparing the Gibbs state \cite{Wiebe2014}. The Gibbs state is first approximated using a classical and efficient mean-field approximation, before being fed into the quantum computer and refined towards the true Gibbs state. The prepared state can then be sampled to calculate the required gradient. The second algorithm performs a similar process, but instead assumes access to the training data in superposition (i.e. through a QRAM - see Section~\ref{sec:QRAM}), which reduces the complexity dependence quadratically on the number of training vectors. For a comparison of deep learning algorithm complexities, see the table in Appendix \ref{app:tablealgorithms}.
 
Wiebe et al. acknowledge that their algorithm is not efficient for all BM configurations, however it is conjectured that these problem BMs will be rare. The improvement of quantum deep learning (QDL) over CDL is the quality of the output as, in the asymptotic limit, it converges to the exact solution. The classical algorithms can only converge to within a certain error, depending on the approximation used. QDL does not provide a speedup over CDL, and so the advantage only comes from the accuracy of the solution. Recent discussions with the authors of the QDL model has highlighted other possible approaches such as using a Heisenberg model instead of an Ising model or, alternatively, using a Bose-Einstein or a Fermi-Dirac distribution instead of a Gibbs distribution. These alternative models may allow for novel machine learning algorithms that don't exist classically.

Recently, an efficient classical equivalent of the QDL algorithm has been found \cite{Wiebe2015}. This suggests QML algorithms may lead to new classical algorithms, stressing the importance of this research even without a working quantum computer.

\subsection{Bayesian Networks} \label{sec:BN}

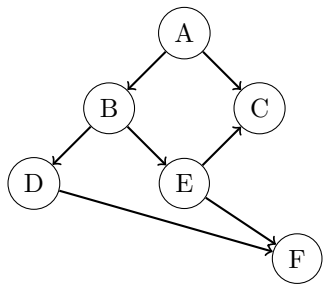
\begin{figure}[h]
\centering
\begin{tikzpicture}

  \node[circle, draw=black](a) at (0,0) {A};
  \node[circle, draw=black](b) at (-1,-1) {B};
  \node[circle, draw=black](c) at (1,-1) {C};
  \node[circle, draw=black](d) at (-2,-2) {D};
  \node[circle, draw=black](e) at (0,-2) {E};
  \node[circle, draw=black](f) at (1.5,-3) {F};
  \draw[->, black, thick] (a) -- (b);
  \draw[->, black, thick] (a) -- (c);
  \draw[->, black, thick] (b) -- (e);
  \draw[->, black, thick] (b) -- (d);
  \draw[->, black, thick] (d) -- (f);
  \draw[->, black, thick] (e) -- (f);
  \draw[->, black, thick] (e) -- (c);
\end{tikzpicture}
\caption{An example of a Bayesian network. Each variable is a node with dependences represented by edges.}\label{fig:BN1}
\end{figure}

A Bayesian network (BN) is a probabilistic directed acyclic graph representing a set of random variables and their dependence on one another (see Figure \ref{fig:BN1}). BNs play an important role in machine learning as they can be used to calculate the probability of a new piece of data being sorted into an existing class by comparison with training data.

Each variable requires a finite set of mutually exclusive (independent) states. A node with a dependent is called a parent node and each connected pair has a set of conditional probabilities defined by their mutual dependence. Each node depends only on its parents and has conditional independence from any node it is not descended from \cite{Nilsson1998}. Using this definition, and taking  $n$ to be the number of nodes in the set of training data, the joint probability of the set of all nodes, $\{X_1,X_2,\cdots X_n\}$, is defined for any graph as

\begin{equation}
P(X_i)=\prod_{i=1}^nP(X_i|\pi_i),
\end{equation}
where $\pi_i$ refers to the set of parents of $X_i$. Any conditional probability between two nodes can then be calculated \cite{Ben-Gal2007}.

An argument for the use of BNs over other methods is that they are able to ``smooth" data models, making all pieces of data usable for training \cite{Heckerman1996}. However, for a BN with $m$ nodes, the number of possible graphs is exponential in $n$; a problem which has been addressed with varying levels of success \cite{Balasubramanian2014, Sclove2011a}. The bulk of the literature on learning with BNs utilises model selection. This is concerned with using a criterion to measure the fit of the network structure to the original data, before applying a heuristic search algorithm to find an equivalence class that does well under these conditions. This is repeated over the space of BN structures. A special case of BNs is the dynamic (time-dependent) hidden Markov model (HMM), in which only outputs are visible and states are hidden. Such models are often used for speech and handwriting recognition, as they can successfully evaluate which sequences of words are the most common \cite{Gales2007}. 

Quantum Bayesian networks (QBNs) and hidden quantum Markov models (HQMMs) have been demonstrated theoretically in several papers, but there is currently no experimental research \cite{Monras2010, Clark2014, Tucci1995}. The format of a HMM lends itself to a smooth transition into the language of open quantum systems. Clark et al. claim that open quantum systems with instantaneous feedback are examples of HQMMs, with the open quantum system providing the internal states and the surrounding bath acting as the ancilla, or external state \cite{Clark2014}. This allows feedback to guide the internal dynamics of the system, thus conforming to the description of an HQMM. 

\subsection{HHL: Solving Linear Systems of Equations} \label{sec:HHLalg}

HHL (named after its discoverers: Harrow, Hassidim and Lloyd) is an algorithm for solving a system of linear equations: given a matrix $A$ and a vector $\mathbf{b}$, find a vector $\mathbf{x}$ such that $A\mathbf{x}=\mathbf{b}$ \cite{Harrow2009, Childs2015}. While this may seem like an esoteric problem unrelated to machine learning, they are in fact intimately related. For example, it is demonstrated in Appendix \ref{app:perceptron} that the HHL algorithm can be employed to give a speedup in perceptron training, that is exponential in the size of the training vectors. It is also the underlying mechanism behind data-fitting procedures such as linear regression and, as such, becomes a workhorse for generic classification problems.

As a more specific example, consider a generic quadratic functional on a vector, $\mathbf{x}$:

\begin{equation}
f[\mathbf{x}]=\mathbf{x}^{\intercal}A\mathbf{x} + \mathbf{b}^{\intercal}\mathbf{x} + c.
\end{equation}
The vector at which this takes the minimum value is found in the usual way, by differentiating with respect to $\mathbf{x}$ and finding where the derivative vanishes. Doing so yields the equation

\begin{equation}
A\mathbf{x} - \mathbf{b} = 0,
\end{equation}
which is the linear equation that is solved by HHL. Therefore, any optimisation problem where the input is a data vector and the function to be optimised over looks (at least locally) quadratic, is really just solving a linear system of equations. HHL, then, may be considered a tool for solving generic optimisation problems in the same sense as classical methods like gradient descent or the conjugate gradient method. In contrast to other machine learning algorithms, the HHL algorithm is completely prescribed in the literature, strict lower bounds are known about its runtime and its caveats can be explicitly stated (for a discussion on these, see Appendix~\ref{app:perceptron} and \cite{Aaronson2015}).

\subsection{Principal Component Analysis}\label{sec:PCA}

\begin{figure}[h]
  \centering
  \includegraphics[width=.8\textwidth]{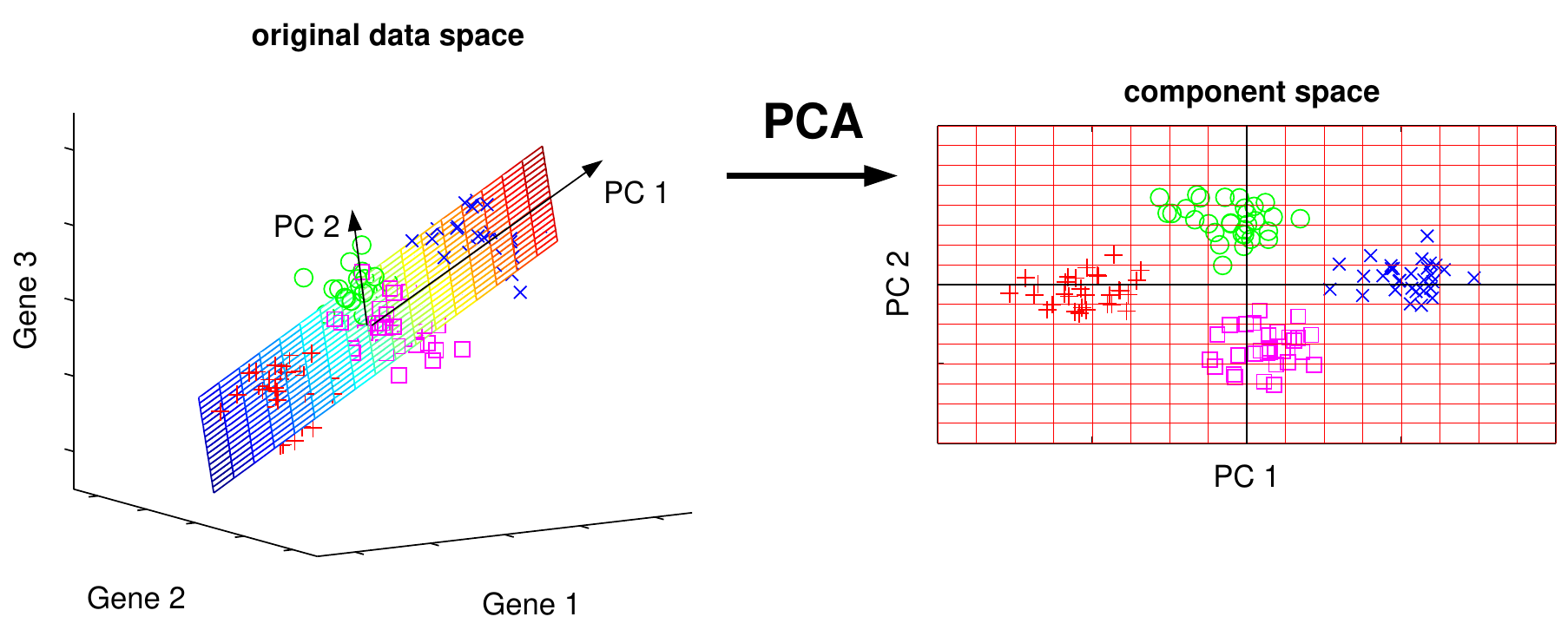}
  \caption{The transformation performed during principal component analysis, on an example dataset. Note that the component space is of lower dimension than the data space but retains most of the distinguishing features of the four groups pictured. Image originally presented in \cite{Scholz2006}. Reprinted with permission from the author.}
  \label{fig:PCA}
\end{figure}

Machine learning data is usually (very) high dimensional, containing redundant or irrelevant information. Thus, machine learning benefits from pre-processing data through statistical procedures such as principal component analysis (PCA). PCA reduces the dimensionality by transforming the data to a new set of uncorrelated variables (the principal components) of which the first few retain most of the variation present in the original dataset (see Figure~\ref{fig:PCA}). The standard way to calculate the principal components boils down to finding the eigenvalues of the data covariance matrix (for more information see reference~\cite{Jolliffe2002}).

Lloyd, Mohseni and Rebentrost recently suggested a quantum version of PCA (QPCA)~\cite{Lloyd2014}. The bulk of the algorithm consists of the ability to generate the exponent of an arbitrary density matrix $\rho$ efficiently. More specifically, given $n$ copies of $\rho$, Lloyd et al. propose a way to apply the unitary operator $e^{- i \rho t}$ to any state $\sigma$ with accuracy $\epsilon = O(t^2/n)$. This is done using repeated infinitesimal applications of the swap operator on $\rho \otimes \sigma$. Using phase estimation, the result is used to generate a state which can be sampled from to attain information on the eigenvectors and eigenvalues of the state $\rho$. The algorithm is most effective when $\rho$ contains a few large eigenvalues and can be represented well by its principal components. In this case, the subspace spanned by the principal components $\rho '$ closely approximates $\rho$, such that $\Vert \rho - P \rho P \Vert \leq \epsilon$, where $P$ is the projector onto $\rho '$. This method of QPCA allows construction of the eigenvectors and eigenvalues of the matrix $\rho$ in time $O(R\mbox{log}(d))$, where $d$ and $R$ are the dimensions of the space spanned by the $\rho$ and $\rho '$ respectively. For low-rank matrices, this is an improvement over the classical algorithm that requires $O(d)$ time. In a machine learning context, if $\rho$ is the covariance matrix of the data, this procedure performs PCA in the desired fashion.

The QPCA algorithm has a number of caveats that need to be covered before one can apply it to machine learning scenarios. For example, to gain a speedup, some of the eigenvalues of $\rho$ need to be large (i.e. $\rho$ needs to be well approximated by $\rho'$). For the case where all eigenvalues are equal and of size $O(1/d)$, the algorithm reduces to scaling in time $O(d)$ which offers no improvement over classical algorithms. Other aspects that need to be considered include the necessity of QRAM and the scaling of the algorithm with the allowed error $\epsilon$. As of yet, it is unclear how these requirements affect the applicability of the algorithm to real scenarios. A useful endeavour would be to perform an analysis for the QPCA algorithm similar to Aaronson's for HHL~\cite{Aaronson2015}. It is not hard to imagine that there is such a dataset that satisfies these caveats, rendering the QPCA algorithm very useful.

\subsection{A Quantum Nearest-Centroid Algorithm for $k$-Means Clustering} \label{sec:kmeans_lloyd}
$k$-means clustering is a popular machine learning algorithm that structures an unlabelled dataset into $k$ classes. $k$-means clustering is an NP-hard problem \cite{Mahajan2009}, but examining methods that reduce the average-case complexity is an open area of research. A popular way of classifying the input vectors is to compare the distance of a new vector with the centroid vector of each class (the latter being calculated from the mean of the vectors already in that class). The class with the shortest distance to the vector is the one to which the vector is classified. We refer to this form of classification sub-routine for $k$-means clustering, as the nearest-centroid algorithm.

\begin{figure}[h!]
\centering
\includegraphics[width=0.40\textwidth]{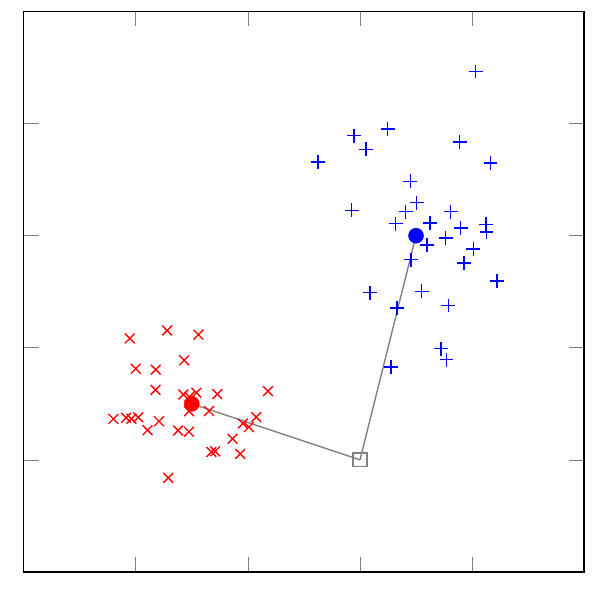}
\caption{A visualisation of the nearest-centroid classification algorithm. Here, a training dataset (with vectors of dimension 2) has already been clustered into two classes. Input data can be classified by comparing the distance in the feature space from the mean position of each cluster. The pluses and crosses represent vector positions of different classes on the feature space, filled circles represent cluster mean positions and the square is a vector that is yet to be classified.}
\end{figure}

 Lloyd et al. have constructed a quantum nearest-centroid algorithm \cite{Lloyd2013}, only classifying vectors after the optimal clustering has been found. They show that the distance between an input vector, $\ket{u}$, and the set of $n$ reference vectors $\{\ket{v^C_i}\}$ of length $m$ in class $C$, can be efficiently calculated to within error $\epsilon$ in $O(\epsilon^{-1}\log n m)$ steps on a quantum computer. The algorithm works by constructing the state
\begin{equation}
\ket{\psi}=\frac{1}{\sqrt{2}}\left(\ket{u}\ket{0}+\frac{1}{\sqrt{n}}\sum_{j=1}^{n}\ket{v^c_j}\ket{j}\right),
\end{equation}
and performing a swap test with the state
\begin{equation}
\ket{\phi}=\frac{1}{\sqrt{Z}}\left(|\mathbf{u}|\ket{0}-\frac{1}{\sqrt{n}}\sum_{j=1}^{n}|\mathbf{v}^c_j|\ket{j}\right),
\end{equation}
where $Z=|\mathbf{u}|^2+(1/n)\sum_j |\mathbf{v}_j|^2$. The distance between the input vector and the weighted average of the vectors in class $C$ is then proportional to the probability of a successful swap test (see Appendix \ref{app:QSVM} for proof). The algorithm is repeated for each class until a desired confidence is reached, with the vector being classified into the class from which it has the shortest distance. 

The complexity arguments on the dependence of $m$ were rigorously confirmed by Lloyd et al. \cite{Rebentrost2014} using the QPCA construction for a support vector machine (SVM) algorithm. This can roughly be thought of as a $k$-means clustering problem with $k$=2. A speedup is obtained due to the classical computation of the required inner products being $O(nm)$\footnote{Aaronson has pointed out that a classical random sampling algorithm can perform the same task in an average run time of $\log(nm)/\epsilon^2$~\cite{Aaronson2015}, compared to the quantum complexity of $O(\epsilon^{-1}\log n m)$. Even in this case, the quantum nearest-centroid algorithm has a quadratic advantage in error scaling.}.

The algorithm has some caveats, in particular it only classifies data without performing the harder task of clustering, and assumes access to a QRAM (see Section~\ref{sec:QRAM}).

In the same paper \cite{Lloyd2013}, Lloyd et al. develop a $k$-means algorithm, including clustering, for implementation on an adiabatic quantum computer. The potential of this algorithm is hard to judge, and is perhaps less promising due to the current focus of the quantum computing field on circuit-based architectures.

\subsection{A Quantum $k$-Nearest Neighbour Algorithm}\label{sec:kNN}

$k$-nearest neighbours ($k$-NN) is a supervised algorithm where test vectors are compared against labelled training vectors. Classification of a test vector is performed by taking a majority vote of the class for the $k$ nearest training vectors. In the case of $k$=1, this algorithm reduces to an equivalent of nearest-centroid. $k$-NN algorithms lend themselves to applications such as handwriting recognition and useful approximations to the traveling salesman problem \cite{Johnson1997}.  

$k$-NN has two subtleties. Firstly, for datasets where a particular class has the majority of the training data points, there is a bias towards classifying into this class. One solution is to weight each classification calculation by the distance of the test vector from the training vector, however this may still yield poor classifications for particularly under-represented classes. Secondly, the distance between each test vector and all training data must be calculated for each classification, which is resource intensive. The goal is to seek an algorithm with a favourable scaling in the number of training data vectors.

An extension of the nearest-centroid algorithm in Section~\ref{sec:kmeans_lloyd} has been developed by Wiebe et al. \cite{Wiebe2014knn}. First, the algorithm prepares a superposition of qubit states with the distance between each training vector and the input vector, using a suitable quantum sub-routine that encodes the distances in the qubit amplitudes. Rather than measuring the state, the amplitudes are transferred onto an ancilla register using coherent amplitude estimation. Grover's search is then used to find the smallest valued register, corresponding to the training vector closest to the test vector. Therefore, the entire classification occurs within the quantum computer, and we can categorise the quantum $k$-NN as an $L_2$ algorithm.

The advantage over Lloyd et al.'s algorithm is that the power of Grover's search has been used to provide a speedup and it provides a full and clear recipe for implementation. Using the same notation as in Section~\ref{sec:kmeans_lloyd}, the time scaling of the quantum $k$-NN algorithm is complex, however it scales as $\tilde{O}(\sqrt{n}\log(n))$ to first order. The dependence on $m$ no longer appears except at higher orders.

The quantum $k$-NN algorithm is not a panacea. There are clearly laid out conditions on the application of quantum $k$-NN, though, such as the dependence on the sparsity of the data.  The classification is decided by majority rule with no weighting and, as such, it is unsuitable for biased datasets. Further analysis would involve finding realistic datasets and the number of qubits required for a proof-of-principle demonstration.

\subsection{Other Notable Algorithms}\label{sec:ONA}

Other algorithms, such as minimum spanning tree \cite{Zahn1971}, divisive clustering, and k-medians \cite{Kaufmann1987}, have notable applications in the field of unsupervised machine learning.
 $k$-medians differs from the $k$-means algorithm only by the fact that a median has to be part of the dataset.

In a quantum regime, the aim is to improve these algorithms using quantum
sub-routines for faster classification, or by a wholesale quantum implementation. Unsupervised CML is a mature field, but unsupervised QML still has a lot of potential for further investigation \cite{Wittek2014,Aimeur2012}.

\begin{figure}[htbp]
  \centering
  \begin{tikzpicture}[scale = .4]
    \tikzset{data/.style={draw,
      shape=circle,fill=blue},}
    \begin{scope}
      \draw(-1,8) node{B};
      \path (0,0) node[data] (p0){} (1,2) node[data] (p1){} (1,1) node[data] (p2){}
      (5,6) node[data] (p3){} (6,7) node[data] (p4){} (5,8) node[data] (p5){} (0,5) node[data]
      (p6){} (1,4) node[data] (p7){} (1.5,5.5) node[data] (p8){}; \draw (p0) -- (p2)
      -- (p1); \draw[red, dashed] (p1) -- (p7); \draw (p7) -- (p6) -- (p8);
      \draw[red, dashed] (p8) -- (p5); \draw (p3) -- (p4) -- (p5);
    \end{scope}
    
    \draw[->] (-4,2.8) -- ++(2,0);

    \begin{scope}[shift={(-11cm, 0cm)}]
      \draw(-1,8) node{A};
      \path (0,0) node[data] (p0){} (1,2) node[data] (p1){} (1,1) node[data] (p2){}
      (5,6) node[data] (p3){} (6,7) node[data] (p4){} (5,8) node[data] (p5){} (0,5) node[data]
      (p6){} (1,4) node[data] (p7){} (1.5,5.5) node[data] (p8){}; 
    \end{scope}
    
    \draw[->] (7,2.8) -- ++(2,0);
    
    \begin{scope} [shift={(11cm, 0cm)}]
      \draw(-1,8) node{C};
      \path (0,0) node[data] (p0){} (1,2) node[data] (p1){} (1,1) node[data] (p2){}
      (5,6) node[data] (p3){} (6,7) node[data] (p4){} (5,8) node[data] (p5){} (0,5) node[data]
      (p6){} (1,4) node[data] (p7){} (1.5,5.5) node[data] (p8){}; \draw (p0) -- (p2)
      -- (p1); \draw (p7) -- (p6) -- (p8);
      \draw (p3) -- (p4) -- (p5);
    \end{scope}
  \end{tikzpicture}
  \caption{a) Representation of samples in the dataset. The minimum spanning tree algorithm tries to find a path visiting every sample once, with minimum distance. b) The minimum spanning tree for the given dataset. Quantum sub-routines can be used to speed up the process of finding such a path. c) Clustering into $k$-subsets can easily be achieved by removing the $k-1$ longest legs in the minimum spanning tree (here $k$=3, and Figure \ref{fig:spantree}b marks the longest legs dashed, in red).}
  \label{fig:spantree}
\end{figure}
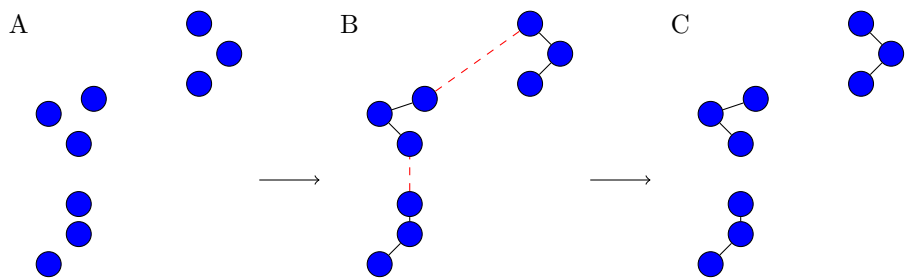

The goal of the minimum spanning tree algorithm is to find the shortest path (the minimum spanning tree) connecting all data points in a feature space (see Figure~\ref{fig:spantree}).

Once the spanning tree has been found, clustering into $k$ classes can be achieved by removing the $k-1$ longest connections. Finding such a path takes $\Omega(N^2)$ time on a classical computer, where $N$ is the number of elements in the dataset. D\"urr et al. show that if the connection matrix of the minimum spanning tree can be given by a quantum oracle, the computational time on a quantum computer can be reduced using Grover's algorithm to $\Theta(N^{3/2})$ \cite{Durr2004}. This algorithm can than be used to find nearest neighbours in a dataset, which is of great importance in graph problems such as the one described above. 

Similarly, divisive clustering and k-medians algorithms can be improved using quantum sub-routines. In divisive clustering, a sub-routine for finding maxima \cite{Durr1996} helps with locating the data points furthest apart within a dataset of similar features. k-medians algorithms can be improved by finding the medians in the dataset via a quantum routine \cite{Aimeur2012}.

\begin{figure}[h]
  \centering
  \includegraphics[width=.8\textwidth]{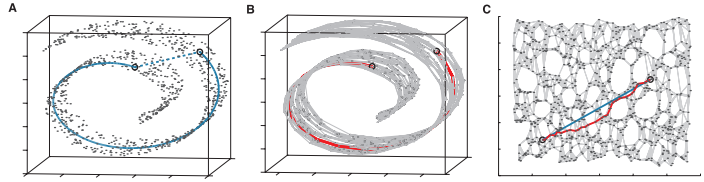}
  \caption{The ``swiss roll'' dataset. a) The euclidean distance (blue dashed line) of two points is shorter than on the manifold, hence they may seem to be more similar then they actually are. b) Geodesic distance on the manifold, calculated via the nearest neighbour graph. c) Unfolded manifold forming the lower dimensional ``true" feature space. The nearest neighbour graph is depicted with gray lines. The geodesic distance calculated via the nearest neighbour graph is shown in red.  Image originally presented in \cite{Tenenbaum2000}. Reprinted with permission from AAAS.}
  \label{fig:swiss-roll}
\end{figure}

Great potential for quantum algorithms also lies in the field of dimensionality reduction algorithms (see Section~\ref{sec:PCA} also). These algorithms attempt to find lower dimensional manifolds, which contain all data points, in a high dimensional feature space. To ``unfold'' these manifolds, for example in the Isomap algorithm \cite{Tenenbaum2000}, it is again important to calculate a graph of nearest neighbours (Figure~\ref{fig:swiss-roll}).
Due to the immense complexity of these algorithms, especially for large datasets, using algorithms as described by D\"urr et al. for dimensionality reduction appears promising.
\section{Experimental implementations}\label{sec:Implementations}

As we are still only in the early stages of developing quantum computers, algorithmic implementations are limited. In this section we outline known experimental results related to QML.

\subsection{Adiabatic Quantum Machine Learning}

%
QML is an obvious avenue of exploration for D-Wave Systems, who are cited by the press as having developed the world's first commercially available quantum computer. The very nature of industry means publically available technical data from D-Wave is limited. However, a number of releases have been made in conjunction with organisations such as Google, discussing QML on an adiabatic quantum computer \cite{Neven2008a,Neven2008b,Neven2009a,Neven2009b,Neven2012,Pudenz2012}. While some papers focus on mapping specific problems to an appropriate D-Wave input format \cite{Neven2008a}, others are more concerned with training binary classifiers \cite{Neven2008b,Neven2009a,Neven2009b,Neven2012}. The former is equivalent to state-preparation without QRAM (see Section~\ref{sec:QRAM}), but one should pay careful attention as to why a large number of different quantum states need not be prepared when handling big data. By deleting entries that fail to demonstrate a certain level of uniqueness and robustness under small image transformations, the dataset experiences a vast reduction in size before any machine learning begins.

In order to select a series of weak classifiers from which a strong classifier can be constructed, D-Wave machines use an algorithm known as \emph{QBoost} \cite{Neven2012}. When compared to the classical \emph{AdaBoost}, an advantage was claimed \cite{Neven2009a}, however the scaling of D-Wave-compatible algorithms compared to optimised classical ones has since been shown to be less than clear-cut \cite{Boixo2014}. Regardless of any quantum effects which may exist within D-Wave's architecture \cite{Lanting2014,Albash2015}, a quantum \emph{speedup} is yet to be demonstrated \cite{Ronnow2014}. Without this, there still exists a possibility that quantum annealing may one day offer a demonstrable advantage in certain situations over conventional computational models, but as of now, \emph{QBoost}'s potential is difficult to quantify. 

\subsection{Implementating a Quantum Support Vector Machine Algorithm on a Photonic Quantum Computer}

Cai et al.~\cite{Cai2015} have implemented a highly simplified version of Lloyd's supervised nearest-centroid algorithm detailed in Section \ref{sec:kmeans_lloyd}. Instead of using $k$-means to cluster training data, a reference vector is chosen for each of their two classes, $\mathbf{v}_A$ and $\mathbf{v}_B$. The task is then to compare whether a new input vector $\mathbf{u}$ is closer to the reference vector of A or B, that is to find the distances
\begin{equation}
D_A=|\mathbf{u}-\mathbf{v}_A|, ~~\text{and}~~D_B=|\mathbf{u}-\mathbf{v}_B|.
\end{equation}
If $D_A < D_B$ then the vector is classified as being in class $A$, and vice versa. As the classifier is deciding between two classes, the quantum computer is performing the classification sub-process in a classical SVM algorithm.

The implemented algorithm computes and stores all vectors classically before inputting them into the quantum computer. Once the swap test has been performed, the distances $D_A$ and $D_B$ are calculated and compared, again classically (see Appendix \ref{app:QSVM} for more details).

\begin{figure}[h!]
\centering
\includegraphics[width=1\textwidth]{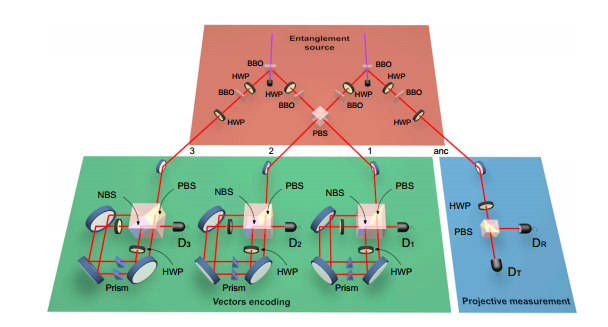}
\caption{The experimental setup used by \cite{Cai2015} to perform the quantum nearest-centroid algorithm. BBO: bismuth-borate crystal; PBS: polarisation beam splitter; HWP: half-wave plate; NBS: normal beam splitter; $\text{D}_i$: detector $i$. Figure originally presented in~\cite{Cai2015}. Reprinted with permission from APS.}
\label{figure:SVMsetup}
\end{figure}

The actual experimental implementation of the quantum sub-process is performed on a small photonic quantum computer, the configuration of which is presented in Figure \ref{figure:SVMsetup}. Two bismuth borate crystals act as pair sources for entangled photons, using spontaneous parametric down-conversion to create a four-qubit entangled state encoded in the photon polarisations. Three of the qubits are sent to Sagnac-like interferometers, where waveplates are used to modify the qubit polarisations such that arbitrary 8-dimensional input and reference vectors can be encoded\footnote{To encode smaller dimensional vectors, a qubit (or two) can be left unmodified such that it is in an equal superposition of horizontal and vertical polarisations, meaning that no information can be gained from the photon's detection and the measurement result can be ignored.}. The fourth qubit acts as an ancilla, and is sent to a polarising beamsplitter (PBS) such that it has a equal chance of being detected in either the reflected or transmitted output mode. All four photons are measured at the same time with four detectors. The probability of the four-fold coincidence $D_3 D_2 D_1 D_R$ and $D_3 D_2 D_1 D_T$ determines the distance between the two vectors. The process can be repeated an arbitrary number of times with the same state preparation settings, to a desired accuracy of $\epsilon$. In the paper, data is (selectively) provided, and the classifier is demonstrated to be prone to errors when the difference between $D_A$ and $D_B$ is small. Whilst the vectors are classified correctly the majority of the time, the experimentally found distances typically differ from the true distances, even after 500 measurements per vector (taking approximately 2 minutes). It is unclear whether this is an inherent problem with the algorithm or an artifact of the system it has been implemented on.

\subsection{Implementing a Quantum Support Vector Machine on a Four-Qubit Quantum Simulator}

\begin{figure}[H]
\centering
\includegraphics[scale=0.5]{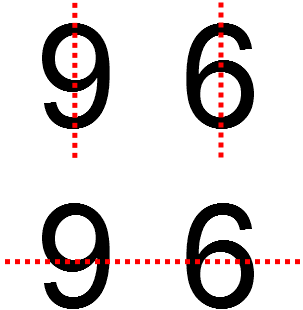}
\caption{Splitting a character in half, either horizontally or vertically, enables it to be classified in a binary fashion. To identify whether a hand-written input is a 6 or a 9, the proportion of the character's constituent pixels which lie on one side of the division are compared with correspondent features from a type-written training set. Based on an image originally presented in~\cite{Zhaokai2014}.}\label{69ratio}
\end{figure}

A recent attempt at implementing quantum machine learning using a liquid-state nuclear magnetic resonance (NMR) processor was carried out by Li et al.~\cite{Li2015}. Their approach focused on solving a simple pattern recognition problem of whether a hand-written number was a 6 or a 9. This kind of task can usually be split into preprocessing, image division, feature extraction and classification. First, an image containing a number of characters will be fed into the computer and transformed to an appropriate input format for the classification algorithm. If necessary, a number of other adjustments can be made at this stage, such as resizing the pixels. Next, the image must be split by character, so each can be categorised separately. The NMR-machine built by Li et al. is only configured to accept inputs which represent single digits, so this step was omitted. Key features of the character are then calculated and stored in a vector. In the case of Li et al., each number was split along the horizontal and vertical axes (Figure~\ref{69ratio}), such that the pixel number ratio across each division could be ascertained. These ratios (one for the horizontal split and one for the vertical) work well as features, since they are heavily dependent on whether the digit is a 6 or a 9. Finally, the features of the input characters are compared with those from a training set. In this case, the training set was constructed from numbers which had been type-written in standard fonts, allowing the machine to determine which class each input belonged to.

\noindent
In order to classify hand-written numbers, Li et al. used a quantum support vector machine. As mentioned in Section~\ref{sec:kmeans_lloyd}, this is simply a more rigorous version of Lloyd's quantum nearest centroid algorithm \cite{Rebentrost2014}.

We define a normal vector, $\mathbf{n}$, as
\begin{equation}
\mathbf{n}=\sum_{i=1}^{m}w_{i}\mathbf{x}_i,
\end{equation}
where $w_{i}$ is the weight of the training vector $x_{i}$.
The machine then identifies an optimal hyperplane (a subspace of one dimension less than the space in which it resides), satisfying the linear equation
\begin{equation}
\mathbf{n}\cdot\mathbf{x}+c=0,
\end{equation}
The optimisation procedure consists of maximising the distance $2 / \left|\mathbf{n}\right|^2$ between the two classes, by solving a linear equation made up of the hyperplane parameters $w_{i}$ and $c$. HHL \cite{Harrow2009} solves linear systems of equations exponentially faster than classical algorithms designed to tackle the same problem. Therefore, it is hoped that reformulating the support vector machine in a quantum environment will also result in a speedup. 

After perfect classification we find that, if $\mathbf{x}_{i}$ corresponds to the number 6,
\begin{equation}
\mathbf{n}\cdot\mathbf{x}_{i}+c\geq 1,
\end{equation}
whereas if it corresponds to the number 9,
\begin{equation}
\mathbf{n}\cdot\mathbf{x}_{i}+c\leq -1.
\end{equation}
As a result, it is possible to determine whether a hand-written digit is a 6 or a 9 simply by evaluating where its feature vector resides with respect to the hyperplane.

\begin{figure}[H]
\centering
\includegraphics[width=0.9\textwidth]{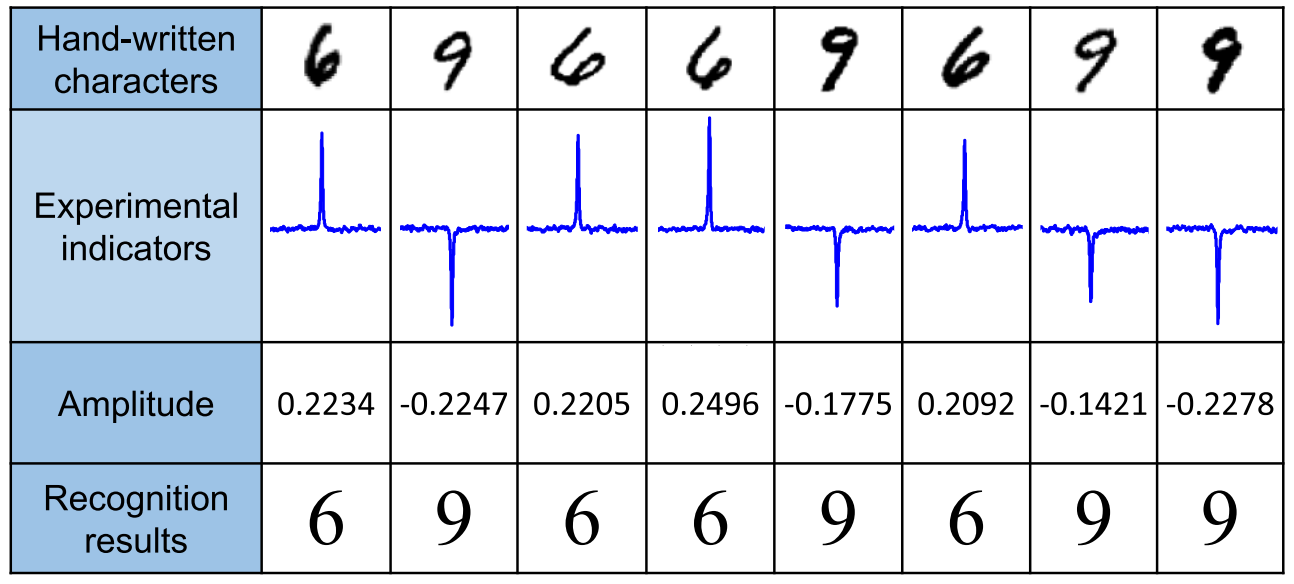}
\caption{The experimental results obtained by Li et al. illustrate that all the hand-written characters have been correctly classified, without ambiguity \cite{Li2015}. Reprinted with permission from APS.}\label{69results}
\end{figure}

The experimental results published by Li et al. are presented in Figure~\ref{69results}. We can see that their machine was able to recognise the hand-written characters across all instances. Unfortunately, it has long been established that quantum entanglement is not present in any physical implementation of liquid-state NMR \cite{Braunstein1999,Menicucci2002}. As such, it is highly likely that the work presented here is only a classical simulation of quantum machine learning.
\section{Challenges}\label{sec:Challenges}
Here we outline some challenges for quantum machine learning that we believe should be taken into account when designing new algorithms and/or architectures.

\subsection{Practical Problems in Quantum Computing}

Whilst great progress has been made in the field of quantum technologies, a general purpose error-corrected quantum computer with a meaningful number of qubits is far from realisation. It is not yet clear how many logical qubits quantum computers require to outperform  classical computers, which are very powerful, but it is thought that QML or quantum simulation may provide the first demonstration of a quantum speedup~\cite{Wittek2014,muller2015promise}. The challenges of producing a `large-scale' quantum computer are well understood.

The obstacles in engineering a quantum computer include ensuring that the qubits remain coherent for the time taken to implement an algorithm, being able to implement gates with $\approx$0.1\% error rates, such that quantum error correction may be performed \cite{fowler2012surface}, and having the qubit implementation be scalabe, such that it admits efficient multiplicative expansions in system size. No current qubit implementation solves all of these problems, though significant progress continues to be made.




\subsection{Some Common Early Pitfalls in Quantum Algorithm Design}

Aside from the problems in constructing a quantum computer, there are a number of nuances to algorithm design. For example, an often overlooked aspect of quantum algorithms is state preparation. Arbitrary state preparation is exponentially hard in the number of qubits for discrete gate sets \cite{Nielsen2011}, providing a bound on the performance of all algorithms, and placing a restriction on the types of states used in initialising an algorithm. Moreover, there exist cases where this addition to the algorithm's complexity is ignored. For instance, a scheme that claims to encode data using the amplitudes available in the exponentially large Hilbert space, is questionable due to this exponential cost (and because of readout issues, which we will discuss in the next paragraph). While realistic schemes must make use of states which are easy to access (assuming discrete gates), this is not necessarily a handicap. For a more detailed examination of state preparation, see Section \ref{sec:QRAM}.

A similar issue to the above is the problem of `readout', which can be the downfall of some attempts to create quantum algorithms. Measurement of a quantum mechanical system results in the collapse of the system's wave function to a single eigenstate of the measurement operator. 
Although it is possible to learn the pre-measurement state using a number of trials exponential in system size, this will kill any potential speedup. Therefore, any algorithm which outputs all of the amplitudes of the final state $\ket{x}$, suffers exponential costs. The only information that can be easily extracted from $\ket{x}$ is a global statistical property, such as the inner product,  $\bra{x}z\rangle$, with some fixed reference state $\ket{z}$, or the location of the dominant amplitudes of $\ket{x}$~\cite{Aaronson2015}. This argues against the existence of a useful quantum algorithm that stores output data in the exponentially large Hilbert space of a quantum state - the data would be impossible to retrieve.

Another potential issue for experiments claiming to have performed QML is the case where a CML algorithm has been implemented using a quantum device. This is simply using a quantum computer to do what a classical computer can achieve equally well, and yields no quantum advantage. Moreover, na\"{i}ve attempts to create a QML algorithm by replacing all the vectors in a CML algorithm with quantum states, are usually unsuccessful at attaining a speed-up. Indeed, these attempts often don't translate feasibly to quantum due to the restrictions of unitary evolution and projective measurement.


\subsection{Quantum Random Access Memory}\label{sec:QRAM}
As stated previously, considering how to encode classical data into quantum states is an important part of any quantum algorithm. In terms of state preparation, information is typically encoded in state amplitudes~\cite{Prakash2014}:
\begin{equation}
\text{Given a vector } \mathbf{x} \in \mathbb{R}^N \text{ stored in memory, create copies of the state } \ket{x} = \frac{1}{\vert x \vert} \sum_i x_i \ket{i} .\nonumber
\end{equation}
QRAM is a theoretical oracle that stores quantum states and allows queries to be made in superposition. The efficiency of the oracle removes any overheads for arbitrary state preparation, which could suppress the claimed quantum speedup of an algorithm. There are number of examples of algorithms that require, or are improved upon, by the application of QRAM (see, for example, references~\cite{Wiebe2014},~\cite{Lloyd2014} and~\cite{Rebentrost2014}).

Before Giovanetti, Lloyd and Maccone (GLM)'s two publications in 2008 \cite{Giovannetti2008a, Giovannetti2008b}, little progress had been made in the development of QRAM architectures. In these papers, GLM generalise classical RAM to the `fanout' scheme, a QRAM architecture based on a bifurcation graph. Each node of the graph is an active quantum gate, and the input qubits must be entangled with $O(N)$ of these when querying superpositions of $N=2^n$ memory cells. Here, $n$ is the number of bits in the address register and the memory cells are placed at the end of the bifurcation graph. Fanout schemes are unrealistic to implement in practice because of decoherence. The fidelity between the desired and actual states addressed with a single faulty gate can drop by a factor of two. GLM proposed the `bucket-brigade' scheme, which replaces the gates with three-level memory elements. Most of these are not required to be active in a single memory call, therefore the number of active elements reduces to $O(\log ^2N)$. Assuming such an architecture, it is possible to use QRAM to generate a quantum state from the $n$-dimensional vector $x$, in time $O(\sqrt n)$. However, by pre-processing the vector, this can be improved to $O(\polylog\left(n\right))$~\cite{Prakash2014}. Suggestions of platforms on which QRAM can be realised include, among others, optical lattices and quantum dots~\cite{Giovannetti2008b}. To the best of the authors' knowledge, there have yet to be any experimental demonstrations of QRAM to date.

Recently, there has been focus on the efficiency of the above implementations in the presence of errors. Arunachalam et al.~\cite{Arunachalam2015} analysed a faulty bucket-brigade QRAM, considering the possibilities of wrong paths caused by bit flip errors. Under this model, the authors argue that the error scaling per gate should perform no worse than $O(2^{-\frac{n}{2}})$ , whereas GLM suggested that a scaling of $O(1/n^2)$ would retain coherence, based on a less rigorous analysis. Further analysis found that oracle errors remove the QRAM preparation speedup so the protocol requires active quantum error correction on every gate to compensate for faulty components. Unfortunately, the active error correction itself removes the speedup, and so it is unlikely that states can be efficiently prepared without noiseless gates. This result also applies to a study that improved the number of time steps per memory call of the bucket-brigade \cite{Hong2012}.

The inclusion of QRAM in QML proposals is troubling, both from a theoretical and an experimental perspective. However ruling out QRAM does not necessarily mean no datasets can be loaded into a quantum state efficiently. If the coefficients to be loaded into a state are given by an explicit formula, it may be possible for a quantum computer to prepare said state independently, without consulting a QRAM. A preprint by Grover and Rudolph \cite{Grover2002} (which was independently discovered by both Zalka \cite{Zalka1998}, and Kaye and Mosca \cite{Kaye2004}) addresses this, by discretising a given function $f(x)$ and sampling at $2^n$ points. This sample can be loaded into a superposition over $n$ qubits efficiently, provided there is an efficient classical algorithm to integrate the function over an arbitrary interval. Whether meaningful datasets fall into this category, or whether there are other categories of functions that can be prepared efficiently, is unclear. However, it is a strong indication that a total dependence on QRAM is not necessary.
\section{Conclusion}\label{sec:conclusion}
Machine learning and quantum information processing are two large, technical fields that must both be well understood before their intersection is explored. To date, the majority of QML research has come from experts in either one of these two fields and, as such, oversights can be made. The most successful advances in QML have come from collaborations between CML and quantum information experts, and this approach is highly encouraging. These early results highlight the promise of QML, where not only are time and space scalings possible but also more accurate classifiers. We would also like to stress that QML research can help guide and advance CML, and therefore it is important to pursue even though we do not yet have the hardware to implement useful instances of QML.

There are several important issues to overcome before the practicality of QML becomes clear. The main problem is efficiently loading large sets of arbitrary vectors into a quantum computer. The solution may come from a physical realisation of QRAM, however as discussed previously, it is currently unclear whether this will be possible.

Despite the potential limitations, research into the way quantum systems can learn is an interesting and stimulating pursuit. Considering what it means for a quantum system to learn could lead to novel quantum algorithms with no classical analogue. The field of CML has already begun to revolutionise society. Any possible addition to this endeavour, and its sometimes unpredictable consequences, should be explored.

\section{Acknowledgments}

We would like to thank Nathan Wiebe for his insightful discussions whilst writing the document. We would also like to acknowledge Seth Lloyd and Patrick Rebentrost for their correspondence regarding QPCA. We are indebted to the doctoral training centre management for their support, and special thanks go to Ashley Montanaro, Peter Turner and Christopher Erven for their feedback and discussion over the course of the unit. The authors acknowledge funding from the EPSRC Centre for Doctoral Training in Quantum Engineering as well as the Defence Science and Technology Laboratory.

\newpage
\pagestyle{empty}
\bibliographystyle{qml2}

\bibliography{QMLlandscape}

\newpage
\pagestyle{fancy}
\begin{appendix}
\appendix
\appendixpage
\section{Quantum Perceptron Model}\label{app:perceptron}

The purpose of these notes is to discuss a quantum algorithm that implements a perceptron, with a worst-case run-time asymptotically better than that which can be expected classically. This algorithm has been independently produced by the quantum engineering cohort.

\subsection{Perceptrons}

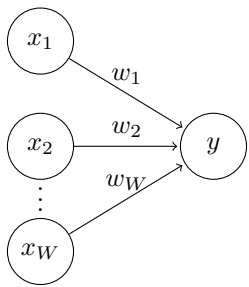
\begin{figure}[H]
\centering
    \begin{tikzpicture}[shorten >=1pt,node distance=2cm,on grid,auto] 
    \node[state] (q_0) {$y$};
    \node[state] (q_1) [left = 2.3cm of q_0] {$x_2$};
    \node[state] (q_2) [below left = 1.4cm and 2.3cm of q_0] {$x_W$};
      \node        (q_dots) [above= 0.8cm of q_2] {$\vdots$}; 
    \node[state] (q_3) [above left = 1.4cm and 2.3cm of q_0] {$x_1$};    

    \path[->,above]
    (q_1) edge node {$w_2$} (q_0)  
    (q_2) edge node {$w_W$} (q_0) 
    (q_3) edge  node {$w_1$} (q_0) 
    ; 


	\end{tikzpicture} 
	\caption{A graphical representation of a perceptron.\label{fig:aPerceptron}}
\end{figure}

A perceptron is a supervised machine learning primitive where the goal is to produce a binary classifier. More specifically, it is a structure that learns given access to some finite set of training data and then uses that knowledge to infer classifications for unlabelled data. Geometrically, a perceptron can be thought of as a very simple neural network: $W$ input nodes, labelled $x_i$ for $0\leq i \leq W$, are each connected to a single output node, $y$, by an edge (see Figure~\ref{fig:aPerceptron}).
These edges themselves have weights, $w_i$. The collection of all input nodes and weights may also be considered vectors, $\mathbf{x}$ and $\mathbf{w}$ respectively. The goal of the perceptron is to classify according to the following expression:

\begin{equation}\label{eq:activation}
y(\mathbf{x},\mathbf{w}) = \left\{
  \begin{array}{lr}
    1 & \text{if } \mathbf{w}\cdot\mathbf{x} + b > 0\\
    0 & \text{otherwise},
  \end{array}
\right.
\end{equation}
where the bias value,$b$, is a regularisation tool that is constant and does not depend on the input. 

Classically, the weights are initialised to take some random values. Then, the training process begins. The training data consists of multiple pairs of input vectors and correct output labels, i.e. the training set $\mathcal{T}$ consists of elements $\mathcal{T} = \{(\mathbf{x}^1,y^1),(\mathbf{x}^2,y^2)\cdots(\mathbf{x}^N,y^N)\}$. A training instance is ``fed'' into the perceptron (i.e. the expression in (\ref{eq:activation}) is calculated) and an output given based on the current set of weights. The learning step is then to update the weights based on the correctness of the classification. The goal is that, with sufficiently large and diverse training instances, the perceptron gains the ability to correctly classify input vectors, $\mathbf{\tilde{x}}$, where the output label is not known beforehand.

The choice of ``activation'' function in Equation~(\ref{eq:activation}), and the structure of the graph itself, may be far more complicated in general than the example provided here. However, this generalisation opens up the field of neural networks which is beyond our scope. Instead, we focus on the simple perceptron example given above.

\subsection{Quantum Perceptrons}
\subsubsection*{Previous Attempts - Data in States}
As discussed in the main body of this document (Section \ref{sec:NN}), there have been a number of efforts at making a viable quantum perceptron~\cite{Altaisky2014,Schuld2014c,Sagheer2013}. Each attempt has generally followed the same problem methodology, namely that of loading up the input vectors $\mathbf{x}_i$ as quantum states, and applying a `weight operator' to achieve a classification $\ket{y_{est}}$. More precisely, the equation describing the perceptron is taken from the classical version of Equation~(\ref{eq:activation}) to
\begin{equation}\label{eq:badQuantumPerceptron}
\ket{y_{est}} = \sum_{j=1}^{W} \hat{w}_j^t\ket{x_j}.
\end{equation}
$\ket{x_j}$ is a quantum state vector that stores features of a single instance of training data, $\hat{w}_j^t$ is the weight operator for each of the training features after the $t^{th}$ training instance, and $j$ sums over all features of a single training instance. The time dependence of  $\hat{w}_j^t$ manifests when there are multiple training data examples, and so each  $\hat{w}_j$ will be updated with the classification of each data sample.

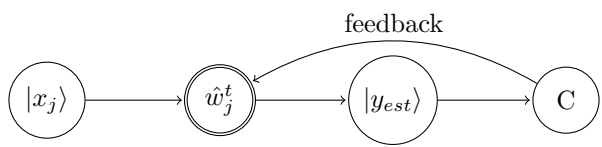
\begin{figure}[ht]
\centering
    \begin{tikzpicture}[shorten >=1pt,node distance=2cm,on grid,auto] 
    \node[state,accepting] (q_0) {$\hat{w}_j^t$};
    \node[state] (q_1) [left = 2.3cm of q_0] {$\ket{x_j}$};
    \node[state] (q_2) [right = 2.3cm of q_0] {$\ket{y_{est}}$};
    \node[state] (q_3) [right = 2.3cm of q_2] {C};

    \path[->,above]
    (q_1) edge node {} (q_0)  
    (q_0) edge node {} (q_2) 
    (q_2) edge node {} (q_3) 
    (q_3) edge [bend right] node {feedback} (q_0) 
    ; 


	\end{tikzpicture} 
	\caption{A flow diagram representing previous approaches to a quantum perceptron.\label{fig:naiveQuantumPerceptron} The node C stands for classification step.}
\end{figure}

This approach is seemingly the most natural progression from Equation~(\ref{eq:activation}), but encounters particular problems when considering implementation of such an algorithm. Figure~\ref{fig:naiveQuantumPerceptron} displays how the method of Equation~(\ref{eq:badQuantumPerceptron}) would work in practice, which is summarised as follows:
\begin{enumerate}
\itemsep0em 
\item Load the features of the first training data instance into the quantum state $\ket{x_j}$.
\item Pass $\ket{x_j}$ through a machine that applies the operator $\hat{w}_j^t$, returning the estimate $\ket{y_{est}}$.
\item Make an appropriate measurement of $\ket{y_{est}}$ (which may be non-trivial, in general) and classify (C) the input training instance;
\item Compare the estimated classification with $\ket{y}$.
\item Feed back the result into the system and alter how the weight operators $\hat{w}_j^t$ act on each feature (the learning step).
\item Repeat steps 1-4 for all instances of training data.
\item After all training data has been used, the operators $\hat{w}_j^t$ are now optimally set to classify unlabelled data.
\end{enumerate}

In a laboratory setting, operations on quantum states are performed using components such as beamsplitters and phase shifters, or by applying some external stimulus like a microwave field. In the case described above, updating how the operator $\hat{w}_j^t$ acts will require the physical turning of knobs or alterations to the experimental setup. An issue with the method proposed is that the learning is still done classically. Step 3 requires measurement of the quantum state and a decision as to how it is classified, before passing that information back into the apparatus that applies $\hat{w}_j^t$, to update how it acts on the system. This is all based on a \textit{classical} update rule. It is difficult to see how a method requiring this level of classical read-out and updating could produce a speedup over a classical technique. 

\subsubsection*{Alternative Approach - Weights in States}
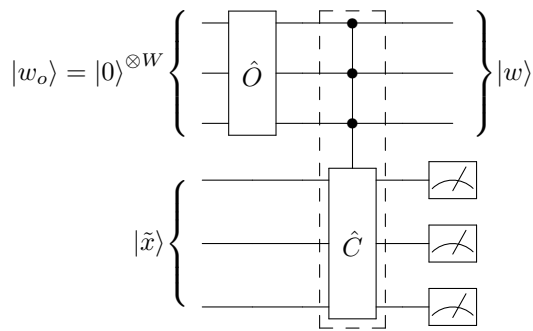
\begin{figure}[ht]
\centerline{
\Qcircuit @C=1em @R=1em {
& & \multigate{2}{\hat{O}} & \qw & \ctrl{1} & \qw & \qw   \\
&\lstick{\ket{w_o} = \ket{0}^{\otimes W} \vast \{ } & \ghost{\hat{O}} & \qw & \ctrl{1} & \qw  & \qw &   \vast \}\ket{w} \\
& & \ghost{\hat{O}} & \qw & \ctrl{1} & \qw  & \qw  \\
& & \qw & \qw & \multigate{2}{\hat{C}} & \qw & \meter \\
& \lstick{\ket{\tilde{x}} \vast \{ } & \qw & \qw & \ghost{\hat{C}} & \qw & \meter \\
& & \qw & \qw & \ghost{\hat{C}} & \qw & \meter \gategroup{1}{5}{6}{5}{.7em}{--}
}
}
\captionof{figure}{A quantum circuit representation of the `weights in states' strategy for a quantum perceptron. The operator $\hat{O}$ performs the training process of the algorithm and the dashed region represents classification of unlabelled data. \label{fig:WiSQuantumCircuit}}\vspace{5mm}
\end{figure}

As an alternative to the `data in states' method, we instead propose that the weights ${w}_j$ be encoded into the state (hence `weights in states'). The method is as follows: 
\begin{enumerate}
\itemsep0em 
\item Initialise a state vector that describes the initial weights for each feature, $\ket{w_o} = \ket{0}^{\otimes W} \in \mathds{C}^{2W}$ (where $W$ is the number of features).
\item Act on this state with an operator $\hat{O}$ which encodes information on the training data you have available.
\item Application of $\hat{O}$ produces a vector of final weights $\ket{w}$ that can be used as a control on the classification of new data. Note that for clarity of discussion, the weights here can only take the value of 0 or 1, and so $\ket{w}$ is a state vector in the computational basis.

The choice of each weight as a computational basis state $\ket{0}$ or $\ket{1}$, is chosen without loss of generality, as the extension involves encoding the binary expansion of the weight in a string of qubits and applying the same operators in an analogous fashion. 
\end{enumerate}
This method can be represented by the quantum circuit displayed in Figure~\ref{fig:WiSQuantumCircuit}. Here we forgo the issues with the previous method by not having to update the operator $\hat{O}$ once it has been initialised. The first problem with such a method however is the definition of suitable operators for $\hat{O}$ and $\hat{C}$. A discussion of possible solutions for these operators is given in the sections below.

\subsubsection*{HHL Algorithm}
\label{sec:HHL}
The HHL algorithm~\cite{Harrow2009, Childs2015} is a quantum algorithm for solving a linear system of equations. That is, it is designed to find a solution vector $\mathbf{x}$ given a matrix $\mathbf{A}$, a vector $\mathbf{b}$, and the equation

\begin{equation}\label{eq:linearsystem}
\mathbf{A}\mathbf{x} = \mathbf{b}.
\end{equation}

Now, consider the approach to perceptrons discussed above. A correctly trained perceptron will have weights such that the correct output is given for all training inputs. That is, it satisfies Equation~(\ref{eq:activation}) for all training vectors. If the sequence of training vectors is labelled by $t$, then the weights must satisfy

\begin{equation}
\mathbf{x}^t \cdot \mathbf{w} + b^t = y^t, \quad t = 1, \cdots, N.
\end{equation}

This set of $N$ equations can be seen as the notation for individual rows of a single matrix equation. Specifically, build a matrix $\mathbf{A}$ with $N$ rows; the elements of the $t^{th}$ row of $\mathbf{A}$ is then given by the elements of $\mathbf{x}^t$. Likewise, define vectors $\mathbf{y}$ and $\mathbf{b}$ such that the $t^{th}$ element is given by $y_t$ and $b_t$, respectively. Then, finding solutions to the set of $N$ equations above, is equivalent to solving the matrix equation

\begin{equation}
\mathbf{A}\mathbf{w} = \mathbf{y}-\mathbf{b} \coloneqq \mathbf{\tilde{y}}.\label{eq:HHL}
\end{equation}

The equation $\mathbf{A}\mathbf{w} = \mathbf{\tilde{y}}$ is exactly the kind of linear system that is solved by the HHL algorithm. Therefore, the operator $\hat{O}$ in Figure~\ref{fig:WiSQuantumCircuit} is given by the matrix $\mathbf{A}$ that must be inverted in order to solve the linear system.

\subsubsection*{Caveats to HHL}
\label{sec:CaveatsHHL}
HHL's ubiquity in quantum computing is hampered by the fact that it has some technical caveats that are often difficult to overcome (see Aaronson~\cite{Aaronson2015}). We will address the three that loom largest: firstly, that HHL requires the matrix $\mathbf{A}$ be \emph{sparse}; secondly, that it takes a quantum state as input, rather than a classical vector; and lastly, that it outputs a quantum state, rather than a classical vector.

The condition on the sparsity of $\mathbf{A}$ is due to the fact that HHL carries out phase estimation as a subroutine. Generally, there are stricter limitations on the simulation of non-sparse Hamiltonians~\cite{Childs2009}. However, the matrix $\mathbf{A}$ is constructed from training data that may or may not be sparse, depending on the problem and context. The only immediate remark is that the field of neural networks with sparse training data still covers a wide range of problems, and is very much an active area of research \cite{Bishop1995}.

It is also necessary for HHL to take a quantum state, here with coefficients drawn from $\mathbf{\tilde{y}}$, as input. Typically, input of this state in a  general context is a non-trivial problem (see the discussion on QRAM in Section \ref{sec:QRAM}). However, for this application, the quantities $y^t$ are just drawn from $\{0,1\}$ (and similarly for the regularisers $b^t$), as the perceptron is a binary classifier. Therefore, preparation of the state $\ket{\tilde{y}}$ just requires preparing a computational basis state. As such, the circuit for this operation is of depth 1 and does not contribute to the complexity of the algorithm.

\begin{figure}[ht]
\centerline{
\Qcircuit @C=1em @R=1em  {
&  & \ctrl{4} & \qw & \qw & \qw & \qw \\
& \lstick{\ket{w} \Bigg \{} & \qw & \ctrl{4} & \qw & \qw & \qw \\
&  & \qw & \qw & \ctrl{4} & \qw & \qw \\
&  &  &  &  &  &  \\
&  & \ctrl{4} & \qw & \qw & \qw & \qw \\
& \lstick{\ket{\tilde{x}} \Bigg \{} & \qw & \ctrl{4} & \qw & \qw & \qw \\
&  & \qw & \qw & \ctrl{4} & \qw & \qw \\
&  &  &  &  &  &  \\
&  & \targ & \qw & \qw & \qw & \meter \\
& \lstick{\ket{0}^{\otimes W} \Vast \{} & \qw & \targ & \qw & \qw & \meter \\
&  & \qw & \qw & \targ & \qw & \meter \\
}
}
\captionof{figure}{A quantum circuit diagram representing the classification process for the `weights in states' method. The circuit above corresponds to the dashed region in Figure~\ref{fig:WiSQuantumCircuit}.}\vspace{5mm}
\label{fig:classification}
\end{figure}
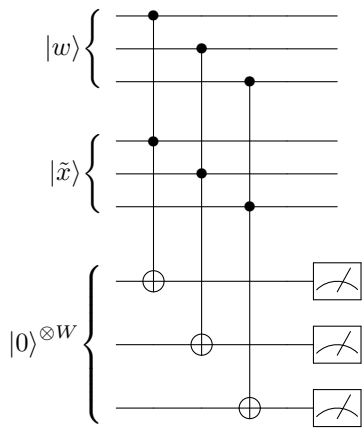

The final caveat that we will address, concerning HHL, is that it outputs a quantum state, rather than a classical vector (i.e. instead of it being a weight vector $\mathbf{w}$, the output is a state $\ket{w}$). In some applications, this is particularly troublesome, as a full reconstruction of the state using quantum tomography would require iterating the algorithm exponentially many times, enough to kill the speedup given by the HHL algorithm. However, this is not a valid concern here. Once the perceptron is trained, and the weights fixed, there is no interest in learning the values of the weights. The goal of the perceptron once it is trained is merely to classify new instances correctly; the weights can remain hidden in a state indefinitely without impacting on its ability to classify new data. A schematic of a circuit suitable for classification is shown in Figure~\ref{fig:classification} and is denoted $\hat{C}$ in Figure~\ref{fig:WiSQuantumCircuit}. Here, the trained weights are encoded in a register $\ket{w}$. The new vector to classify is encoded in a state $\ket{\tilde{x}}$, and there is also a scratchpad of ancillas initialised in the computational basis. The classification process then consists of Toffoli gates, controlled on a weight and an element of the data vector. From Equation~(\ref{eq:activation}), the perceptron classifies as 1 if the dot product between $\mathbf{w}$ and $\mathbf{x}$ is non-zero (ignoring the regularisation term). Therefore as soon as a single element product $x_j w_j$ contributes a positive number, the output should always be 1. Otherwise, the output should be 0. In the circuit, these single element contributions are calculated by measuring the ancillas in the computational basis. If any one of the outputs corresponds to the eigenvalue of $\ket{1}$, the state should be classified as 1. If all of the outputs correspond to the eigenvalue of $\ket{0}$, then the state should be classified as 0. This classification process is non-destructive on the weights and so they can be used repeatedly. As new data and weights are encoded in states in the computational basis, there is no reason why this classification scheme could not be performed equally well classically. However, less na\"{i}ve classification schemes may differ markedly in the quantum and classical cases.

\subsubsection*{Quantum Speedup}

The quantum speedup for this perceptron training algorithm is inherited from the quantum speedup of HHL. Specifically, HHL runs in time

\begin{equation}
O\left(\log{(N)}\cdot\log{\left(\frac{1}{\varepsilon}\right)}\right),
\end{equation}
where $\varepsilon$ is a measure of error in the weight vector~\cite{Childs2015}. Conversely, the best possible classical iterative algorithm for this problem is the conjugate gradient method, which has a runtime of

\begin{equation}
O\left(N \cdot \log{\left(\frac{1}{\varepsilon}\right)}\right).
\end{equation}

HHL offers an exponential improvement in $N$. In addition, it scales similarly in $\varepsilon$ as the best possible classical algorithm because the output of the training stage is a quantum state, rather than some classical data. Both the quantum and classical classification stages run linearly in $W$, giving total runtimes of

\begin{equation}
O\left(W + \log{(N)}\cdot\log{\left(\frac{1}{\varepsilon}\right)}\right)
\end{equation}
quantumly and

\begin{equation}
O\left(W + N \cdot \log{\left(\frac{1}{\varepsilon}\right)}\right)
\end{equation}
classically. Therefore, provided that the number of training instances $N$ is much larger than the number of weights, $W$, we can expect a much better performance from the quantum algorithm. Practically, this is almost always the case. In fact, exponentially more training instances than weights already gives an exponential speedup for the quantum algorithm, even for the na\"{i}ve classification scheme presented here.

\subsection{Discussion}
The construction above assumes that the weights of each feature can only take the value of 0 or 1. The method can be generalised to non-binary weights by encoding a single weight in multiple qubits. This will produce an extra qubit overhead depending on the precision of your weights but should only contribute a logarithmic factor in precision to the total run time of the algorithm. There are a number of other simplifications we have used here which may limit the usefulness of the quantum perceptron. Specifically, we have assumed that there is a solution to the set of linear equations presented in Equation~\ref{eq:HHL}, i.e. there is a combination of weights that classifies all the training data correctly. This is often not the case and it is unclear how the applicability of the algorithm will be affected.

As mentioned previously, perceptrons are a single example of a much larger class of models that are used to classify datasets. Neural networks are to some extent multilayered perceptrons, and are ubiquitous in machine learning \cite{Bishop1995}. Extending the above discussion to neural networks or more complicated perceptron models may produce results that can be widely applied to many classical problems.

\newpage
\section{Probabilistic Distance in the Quantum SVM}\label{app:QSVM}

In this section, we provide a short proof that the distance between two vectors on a quantum computer can be calculated using a probabilistic technique, as utilised in the $k$-nearest centroid algorithm of Section \ref{sec:kmeans_lloyd}. 

Instead of considering the general nearest centroid algorithm, we restrict ourselves to the quantum SVM algorithm. The task is then to compare whether a new input vector $\mathbf{u}$ is closer to the reference vector of A or B, i.e. to find the distances
\begin{equation}
D_A=|\mathbf{u}-\mathbf{v}_A|, ~~\text{and}~~D_B=|\mathbf{u}-\mathbf{v}_B|.
\end{equation}
If $D_A < D_B$ then the vector is classified as being in set $A$, and vice versa. To find these distances on a quantum computer, we first represent the vectors in ket notation:
\begin{equation}
\ket{u}=\frac{\mathbf{u}}{|\mathbf{u}|},~~\text{and}~~\ket{v_i}=\frac{\mathbf{v}_i}{|\mathbf{v}_i|},
\end{equation}
where $i\in\{A,B\}$. These vectors are stored classically, along with their norms which must be calculated. The quantum algorithm is as follows:
\begin{enumerate}
\item{Choose $i=A$. Create a superposition of the states $\ket{u}$ and $\ket{v_i}$, entangled with an ancilla qubit such that
\begin{equation}
\ket{\psi}=\frac{1}{\sqrt{2}}(\ket{0}\ket{u}+\ket{1}\ket{v_i}).
\end{equation}}
\item{Project the ancilla in the prepared state onto
\begin{equation}
\ket{\phi}=\frac{1}{\sqrt{Z}}(|\mathbf{u}|\ket{0}-|\mathbf{v}_i|\ket{1}),
\end{equation}
where $Z=|\mathbf{u}|^2+|\mathbf{v}_i|^2$. The probability of success, $p$, of this projection is given by $|\bra{\phi}\psi\rangle|^2$ where $p$ can be determined to accuracy $\epsilon$ with $O(p(1-p)/\epsilon^2)$ iterations of steps 1 and 2~\cite{Rebentrost2014}.}
\item{\emph{Lemma 1:} The distance can now be calculated classically from
\begin{equation}
D_i=\sqrt{2pZ}.
\end{equation}}
\item{The process is then repeated from step 1, for $i=B$. The two calculated distances are compared classically such that if $D_A-D_B<0$ then $\ket{u}$ is classified as $A$ and if $D_A-D_B>0$, $\ket{u}$ is classified as $B$.
}
\end{enumerate}
The following proof serves to convince the reader the algorithm is indeed measuring the distance between the vectors $\mathbf{u}$ and $\mathbf{v}_i$:
\newline \\
\emph{Proof of Lemma 1:} The probability of the ancilla qubit being in state $\ket{\phi}$ is given by
\begin{equation}
p=| P\ket{\psi}|^2,
\end{equation}
where we have chosen the projector $P$ to be $\ket{\phi}\bra{\phi}\otimes\mathbb{I}$. If we drop the identity term and keep in mind we are acting only on the ancilla, we find
\begin{equation}
p=| P\ket{\psi}|^2=| \ket{\phi}\bra{\phi}\psi\rangle|^2=|\bra{\phi}\psi\rangle|^2.
\end{equation}
By substituting in explicit forms of our states, this becomes
\begin{align}
p&=\frac{1}{2Z}|(|\mathbf{u}|\ket{u}-|\mathbf{v}_i|\ket{v_i})|^2 \nonumber\\
&= \frac{1}{2Z} (|\mathbf{u}|^2 + |\mathbf{v}_i|^2 - |\mathbf{u}| |\mathbf{v}_i|( \bra{u}v_i\rangle+\bra{v_i}u\rangle)) \nonumber\\ 
&= \frac{1}{2Z} (|\mathbf{u}|^2 + |\mathbf{v}_i|^2 - 2 \mathbf{u}\cdot\mathbf{v}_i).
\end{align}
Making the identification of the Euclidean distance for any two vectors $\mathbf{x}$ and $\mathbf{y}$
\begin{equation}
D\equiv|\mathbf{x}-\mathbf{y}|=\sqrt{|\mathbf{x}|^2 + |\mathbf{y}|^2 - 2 \mathbf{x}\cdot\mathbf{y}},
\end{equation}
the result follows
\begin{equation}
p= \frac{1}{2Z} D_i^2 ~~\Rightarrow~~D_i=\sqrt{2pZ}.
\end{equation}
\hfill$\square$

\newpage

\section{Table of Quantum Algorithms}
\label{app:tablealgorithms}
The table below summarises the majority of algorithms discussed in the main body of the text.  Where possible, we include the advantage the quantum algorithm gains over its classical counterpart and any conditions required for the speedup to be maintained. The algorithms are listed in the order that they appear in the main document.
\newpage

\renewcommand{\arraystretch}{1.6}
\thispagestyle{empty}
\newgeometry{left=1cm,bottom=1cm}
\begin{landscape}

\begin{table*}[t!]
\scriptsize{
\centering

\begin{tabular}{|m{2.5cm}||m{3.6cm}|m{2cm}|m{2.5cm}|m{3.4cm}|m{3.3cm}|m{5.4cm}|  }
\hline
\bf{Algorithm}										& \bf{Quantum Time Scaling} 	& \bf{Quantum Space Scaling} & \bf{Classical Time\newline Scaling} & \bf{Definition of Terms} & \bf{Quantum Speedup?} & \bf{Comments}  \\ 
\hhline{|=|=|=|=|=|=|=|} 

Quantum Dot-\newline based Artificial\newline Neural Network~\cite{Behrman1996,Behrman2000, Behrman2002, Behrman1999, Behrman2013, Behrman2008}.	& No rigorous bounds. & - & - & - & Not available. & Might provide spatial benefits because of capability to create temporal neural network, but no rigorous analysis found in papers.   \\ \hline  

Superposition-based Learning\newline Algorithm\newline (WNN) \cite{DaSilva2012a}.	 & $O(\text{poly}(N_p))$ & - & - & $N_p$: No.  training patterns. & Unknown. & Assuming pyramidal QRAM network, and two inputs to every QRAM node.  \\ \hline  

Associative\newline Memory using\newline Stochastic\newline Quantum Walk \cite{Schuld2014a}.			& No rigorous bounds. & - & - & $\kappa$: Coherent weights.\newline $\gamma$: Decoherent  weights. & Not available. & Authors identify dependence of runtime on $\kappa$ and $\gamma$ with no detailed analysis offered.    \\ \hline  

Probabilistic\newline Quantum\newline Memories  \cite{Trugenberger2001}.						& No rigorous bounds.		 & - & - & - & Not available. & Specifically notes no discussion of ``possible quantum speedup" as ``[...] the main point of the present Letter is the exponential storage capacity with retrieval of noisy inputs". \\ \hline  

Quantum Deep \newline Learning (Gibbs \newline Sampling) \cite{Wiebe2014}.					& $\tilde{O}(N E \sqrt{\kappa})$ & $O(n_h + n_v +\newline \hphantom{O( } \log\epsilon^{-1})$ & $\tilde{O}(N L E k)$\newline $O(NE \kappa)$ \newline \newline$L$: No. layers.\newline$k$: No. sample sets.   & $N$: No.  training vectors. \newline $E$: No.  edges.\newline  $\kappa$: Scaling factor.\newline  $x$: Training vector.\newline $n_h$: No.  hidden nodes.\newline $n_v$: No.  visible nodes.  & Asymptotic advantage.  &The quantum advantage lies in the found solution being `exact' up to $\epsilon$. The classical algorithm converges to an approximate solution.  \\ \hline  

Quantum Deep \newline Learning \newline(Amplitude \newline Estimation) \cite{Wiebe2014}.  				& $\tilde{O}(\sqrt{ N} E^2 \sqrt{\kappa} )$ & $O(n_h + n_v +\newline \hphantom{O( } \log\epsilon^{-1})$ &$\tilde{O}(N L E k)$\newline $O(NE \kappa)$ \newline \newline$L$: No. layers.\newline$k$: No. sample sets.   &  $N$: No.  training vectors. \newline $E$: No.  edges.\newline  $\kappa$: Scaling factor.\newline  $x$: Training vector.\newline $n_h$: No.  hidden nodes.\newline $n_v$: No.  visible nodes.   &  Quadratic advantage in number of training vectors.   & Assumes access to training data in quantum oracle. The dependence on $E$ can be reduced quadratically in some cases.  \\ \hline  

Quantum Deep \newline Learning \newline(Amplitude \newline Estimation \newline and QRAM). \cite{Wiebe2014}		 & $\tilde{O}(\sqrt{N} E^2 \sqrt{\kappa})$ & $O(N+n_h + n_v +\newline \hphantom{O( }\log\epsilon^{-1})$ & $\tilde{O}(N L E k)$\newline $O(NE \kappa)$ \newline \newline$L$: No. layers.\newline$k$: No. sample sets.  &   $N$: No.  training vectors. \newline $E$: No.  edges.\newline  $\kappa$: Scaling factor.\newline  $x$: Training vector.\newline $n_h$: No.  hidden nodes.\newline $n_v$: No.  visible nodes.  & Same as Quantum Deep Learning (Amplitude Estimation). & Assumes QRAM allows simultaneous operations on different qubits at unit cost.  \\ \hline 

Linear Systems\newline Solving (HHL)\newline \cite{Harrow2009, Childs2015}.							 &  $\hphantom{O}$Quantum state output:\newline$O( \text{poly} \log(N), \text{poly} \log(\epsilon^{-1}))$\newline $\hphantom{O}\hphantom{O}$Classical output:\newline$\hphantom{O}O( \text{poly} \log N, poly(\epsilon^{-1}))$ & $O(\log N)$ & $O(\text{poly}(N) \log(\epsilon^{-1}))$ & $N$: Dimension of the system.\newline$\epsilon$: Error.  & Exponential if the desired output is a quantum state. Complicated otherwise. & As lots of classes of problems can be reduced to linear system solving, the quantum advantage is heavily dependent on context.  \\ \hline 

Quantum Principal\newline Component\newline Analysis \cite{Lloyd2014}.												& $O(\log{d})$ & - & O(d) &  $d$: Dimensions of\newline \hphantom{$d$: }data space. & Exponential in $d$. & Speedup only valid for spaces dominated by few principal components. Algorithm requires QRAM. \\ \hline  

Quantum Nearest\newline Centroid \newline(sub-routine\newline in k-means) \cite{Lloyd2013, Rebentrost2014}.	 & $O(\epsilon^{-1} \log nm)$ & - & $O(nm)$ & $n$: No.  training vectors.\newline $m$: Length of vectors.  & Exponential speedup, however average classical runtime can be $\log(nm)/(\epsilon^2)$.  & Assumes all vectors and amplitudes stored in QRAM, otherwise speedup vanishes. Quantum Support Vector Machine is Quantum Nearest Centroid with two clusters. \\ \hline  

$k$-nearest\newline Neighbours \cite{Wiebe2014knn}.						& $\tilde{O}( \sqrt{n} \log{n})$ (first order) & - & $O(nm)$ & $n$: No.  training vectors.  & Quantum advantage for high-dimensional vector spaces. & Reduces to nearest centroid for $k$=1.  \\ \hline  
 
Minimum\newline Spanning Tree~\cite{Durr2004}.& $\Theta(N^{3/2})$  & - & $\Omega(N^2)$ & $N$: No.  points in dataset.& Polynomial in the sub-routine used to find a minimum spanning tree. & Matrix model.   \\ \hline  

Quantum\newline Perceptron\newline (Data in States)	\cite{Altaisky2014,Schuld2014c,Sagheer2013}.				 & No rigorous bounds. & - & $O(W+N\log(\epsilon^{-1}))$ & $W$: No. of weights.\newline $N$: Size of training dataset.\newline $\epsilon$: Allowed error. & Not available. & Producing a speedup appears to be difficult using these algorithms (see Appendix~\ref{app:perceptron}, from which the classical bound has been taken). \\ \hline

Quantum\newline Perceptron\newline (Weights in\newline States) [Appx. \ref{app:perceptron}]. 					 & $O(W+\log(N)\log(\epsilon^{-1}))$ & - & $O(W+N\log(\epsilon^{-1}))$ & $W$: No. of weights.\newline $N$: Size of training dataset.\newline $\epsilon$: Allowed error. & Exponential in $N$. &  - \\ \hline

\end{tabular}
}
\end{table*}
\label{table:algorithms}
\end{landscape}
\restoregeometry

\end{appendix}

\end{document}